\documentclass[floatfix,prd,10pt,aps]{revtex4-2}
\usepackage{graphicx,epstopdf}
\usepackage[caption=false]{subfig}
\usepackage{appendix}
\usepackage[T1]{fontenc}
\usepackage{lmodern}
\usepackage[dvipsnames,x11names]{xcolor}
\definecolor{linkc}{RGB}{0,70,140}
\definecolor{citec}{RGB}{170,80,0}
\usepackage[colorlinks=true,linkcolor=linkc,citecolor=citec,urlcolor=linkc]{hyperref} 
\usepackage[sort&compress]{natbib}
\usepackage{morefloats}
\usepackage{amsmath}
\usepackage{amssymb}
\usepackage{amsfonts}
\usepackage{mathtools}
\usepackage{bbold}
\usepackage{multirow}
\usepackage[normalem]{ulem}
\usepackage{physics}
\usepackage{colortbl}
\usepackage{placeins}
\usepackage{booktabs}
\usepackage{slashed}
\usepackage{orcidlink}

\newcommand{\barm}{\bar{m}}

\def\pslash{p\!\!\!\slash }

\def\xslash{x\!\!\!\slash }

\hypersetup{pdfencoding=auto,psdextra}

\begin{document}

\title{Heavy-quark spin symmetry and Pauli exclusion in pentaquarks with three to five heavy quarks}

\author{Ula\c{s} \"{O}zdem\orcidlink{0000-0002-1907-2894}}
\email{ulasozdem@aydin.edu.tr}
\affiliation{Health Services Vocational School of Higher Education, Istanbul Aydin University, Sefakoy-Kucukcekmece, 34295 Istanbul, T\"{u}rkiye}


\begin{abstract}
We study negative-parity hidden-flavor pentaquarks with three to five heavy quarks, $QQ\bar{Q}qq$, $QQQ\bar{Q}q$, and $QQQQ\bar{Q}$ with $Q=c,b$ and $q=u,d$, in QCD sum rules, using the complete set of S-wave baryon--meson currents with $J^P=1/2^-$, $3/2^-$, and $5/2^-$ and the operator product expansion up to dimension 10, 9, and 6. We focus on the splittings between states of the same baryon--meson configuration, which, unlike the absolute masses, are only weakly affected by the choice of the heavy-quark mass scheme and by a finite width. Configurations whose light degrees of freedom have a unique spin are nearly degenerate or show splittings that decrease from charm to bottom by a factor compatible with $\bar m_c/\bar m_b$, whereas configurations with two light quarks in different hadrons show splittings that are not reduced, as expected from heavy-quark spin symmetry. A decomposition of the correlators into factorizable two-hadron and connected quark-exchange contributions shows that the correlators are dominated by the two-hadron part and that this pattern holds separately for both parts: the quark-exchange contribution decreases from charm to bottom when a heavy quark is exchanged and is not reduced when a light quark is exchanged. In the fully heavy systems the $1/2^-$ and $3/2^-$ states are degenerate within about $20$~MeV, while the $5/2^-$ state lies $126$--$139$~MeV ($cccc\bar c$) and $183$--$228$~MeV ($bbbb\bar b$) above the $3/2^-$ state. This splitting is generated by the exchange terms between the identical heavy quarks and does not decrease from charm to bottom, which points to the Pauli principle rather than to a hyperfine effect. Currents whose baryon and meson share no quark flavor are exactly factorizable at this order and cannot distinguish bound states from two-hadron states.
\end{abstract}

\maketitle

\section{Introduction}\label{sec:intro}

The spectroscopy of hadrons containing heavy quarks has changed considerably since the observation of the $X(3872)$ by the Belle Collaboration~\cite{Belle:2003nnu}. A large number of charmonium-like and bottomonium-like states that do not fit into the conventional quark-antiquark and three-quark schemes have been reported since then~\cite{Esposito:2014rxa,Esposito:2016noz,Olsen:2017bmm,Lebed:2016hpi,Nielsen:2009uh,Brambilla:2019esw,Agaev:2020zad,Chen:2016qju,Ali:2017jda,Guo:2017jvc,Liu:2019zoy,Yang:2020atz,Dong:2021juy,Dong:2021bvy,Chen:2022asf,Meng:2022ozq,Wang:2025sic}, including the doubly charmed tetraquark $T_{cc}^+$~\cite{LHCb:2021vvq,LHCb:2021auc}. In the baryon sector, the LHCb Collaboration observed hidden-charm pentaquark candidates in the $J/\psi p$ spectrum of $\Lambda_b^0\to J/\psi pK^-$ decays~\cite{LHCb:2015yax,LHCb:2016lve,LHCb:2019kea} and their strange partners in the $J/\psi\Lambda$ spectrum~\cite{LHCb:2020jpq,LHCb:2022ogu}. The narrow $P_c(4312)$, $P_c(4440)$, and $P_c(4457)$ states lie close to the $\Sigma_c\bar D$ and $\Sigma_c\bar D^*$ thresholds, which has made the internal structure of these states one of the central questions of hadron physics.

The proximity to the thresholds has led to their interpretation as hadronic molecules of a charmed baryon and an anticharmed meson~\cite{Chen:2015loa,Roca:2015dva}, while compact diquark--diquark--antiquark and diquark--triquark configurations have been proposed as well~\cite{Maiani:2015vwa,Lebed:2015tna}. In QCD sum rules, the observed states have been studied with molecular and compact currents~\cite{Chen:2015moa,Wang:2015epa,Azizi:2016dhy}. A feature that distinguishes the molecular picture is the role of heavy-quark spin symmetry: since the interaction does not depend on the spin of the charm quarks in the heavy-quark limit, the $\Sigma_c^{(*)}\bar D^{(*)}$ channels form a multiplet whose members are related to each other~\cite{Liu:2019tjn,Du:2021fmf,Neubert:1993mb}. The pattern of masses within such a multiplet is therefore a sensitive probe of the spin-dependent interactions and of the heavy-quark limit.

The discovery of the doubly charmed baryon $\Xi_{cc}^{++}$~\cite{LHCb:2017iph,LHCb:2018pcs} and of structures in the di-$J/\psi$ mass spectrum~\cite{LHCb:2020bwg,ATLAS:2023bft,CMS:2023owd}, which have been interpreted as fully charmed tetraquarks~\cite{Chao:2020dml,Bedolla:2019zwg,Lu:2020cns,Wang:2021kfv,liu:2020eha,Wu:2024euj}, has shown that hadrons with several heavy valence quarks are accessible at present facilities. The triply heavy baryons $\Omega_{ccc}$ and $\Omega_{bbb}$, which enter the lowest thresholds of the systems with four and five heavy quarks, have been predicted in lattice QCD~\cite{Brown:2014ena} and in quark models~\cite{Liu:2019vtx}, and dibaryons with maximal charm and beauty have been studied on the lattice~\cite{Lyu:2021qsh,Mathur:2022ovu}. These developments have extended the interest to pentaquarks with three or more heavy quarks. Triply heavy pentaquarks with the configuration $qqQQ\bar Q$, which corresponds to our systems with $N_h=3$, have been analyzed phenomenologically~\cite{Sharma:2024pfi}, and hadronic molecules with four charm or beauty quarks have been discussed in Ref.~\cite{Liu:2024pio}. The fully heavy pentaquarks $cccc\bar c$ and $bbbb\bar b$ have been addressed in QCD sum rules~\cite{Zhang:2020vpz,Wang:2021xao,Azizi:2024ito}, in the chromomagnetic interaction model~\cite{An:2020jix}, in constituent and potential quark models~\cite{An:2022fvs,Yan:2021glh,Yang:2022bfu,Liang:2024met}, in the MIT bag model~\cite{Zhang:2023hmg}, with diffusion Monte Carlo~\cite{Gordillo:2024sem,Gordillo:2024blx} and Hartree--Fock methods~\cite{Alonso-Valero:2024jim}, and with mass formulas~\cite{Rashmi:2024ako,Sharma:2025grd,Rostami:2026jyz}; see also Refs.~\cite{Gordillo:2023tnz,Jing:2025iqs}. Fully heavy pentaquarks with both charm and bottom quarks have been analyzed in Refs.~\cite{Yan:2023kqf,Azizi:2025fmx,Azizi:2026oym}. The predicted masses scatter over several hundred MeV for $cccc\bar c$ and over several GeV for $bbbb\bar b$. In compact quark models with all quarks in a common S-wave orbital, the Pauli principle for the four identical heavy quarks admits only $J^P=1/2^-$ and $3/2^-$ states~\cite{An:2020jix,Zhang:2023hmg,Liang:2024met}, whereas cluster and baryon--meson configurations also allow $5/2^-$~\cite{Yang:2022bfu,Alonso-Valero:2024jim}.

These studies mostly address a single quark content, and the sum rule analyses typically a single current and spin-parity. How the spin structure of hidden-charm pentaquarks evolves when light quarks are successively replaced by charm quarks has, to our knowledge, not been examined within one framework. Two opposite tendencies can be anticipated. On the one hand, every additional charm quark reduces the number of light degrees of freedom and suppresses spin-dependent interactions by $1/m_c$, which favors nearly degenerate multiplets. On the other hand, the number of identical charm quarks grows, so that the Pauli principle increasingly restricts the allowed spin couplings and may remove members of a multiplet from the S-wave configuration. The relative importance of these effects can only be assessed if the same method and the same criteria are applied along a sequence of systems. A direct test is provided by the replacement $c\to b$: in the heavy-quark limit the total spin of the light degrees of freedom is conserved separately from the heavy-quark spins, so that splittings within a configuration whose light degrees of freedom have a unique spin should decrease approximately as $1/m_Q$, whereas contributions that do not originate from the heavy-quark spins should not.

QCD sumr ules~\cite{Shifman:1978bx,Shifman:1978by,Ioffe:1981kw,Reinders:1984sr,Colangelo:2000dp, Narison:2002woh, Nielsen:2009uh} are well suited for this purpose, since the interpolating currents can be chosen to have a definite baryon--meson structure and a definite spin, and since the full matrix of correlators between currents of the same spin-parity gives access to the mixing and the overlap of different configurations. In this work we study the chains
\begin{equation}
QQ\bar{Q}qq\;\longrightarrow\; QQQ\bar{Q}q\;\longrightarrow\; QQQQ\bar{Q},\qquad Q=c,\,b,
\label{eq:chain}
\end{equation}
with $N_h=3$, 4, and 5 and $q=u,d$, in which one light quark is replaced by a heavy quark at each step while the systems keep hidden heavy flavor. For each system we construct the complete set of S-wave baryon--meson currents with $J^P=1/2^-$, $3/2^-$, and $5/2^-$ and compute the full correlator matrix in each channel, with the operator product expansion including nonperturbative contributions up to dimension 10, 9, and 6 for $N_h=3$, 4, and 5, respectively. The sum rules are projected onto negative parity and analyzed with common criteria and a self-consistent continuum threshold. For the charm systems we determine the masses and residues of the lowest states, examine with overlap ratios which currents couple to the same states, and analyze the spin content of the currents in the nonrelativistic limit. For both flavors we determine the splittings within each baryon--meson configuration and compare them between charm and bottom, and we decompose the correlators into factorizable two-hadron and connected quark-exchange contributions.

The paper is organized as follows. Section~\ref{sec:formalism} introduces the interpolating currents and the sum rules. The numerical analysis and the criteria for the working regions are described in Sec.~\ref{sec:numerics}. Section~\ref{sec:results} presents and discusses the results for the charm and bottom systems, the spin multiplets and their dependence on the heavy-quark mass, and the comparison with other approaches, and Sec.~\ref{sec:summary} contains our conclusions.

\section{Theoretical framework}\label{sec:formalism}

\subsection{Interpolating currents}\label{sec:currents}

The currents are constructed as products of color-singlet baryon and meson operators. For the baryon part we use
\begin{align}
B &= \epsilon_{abc}\left(q_{1a}^{T} C\gamma_{\alpha} q_{2b}\right)\gamma^{\alpha}\gamma_5\, q_{3c}, \label{eq:B}\\
B^{*}_{\mu} &= \epsilon_{abc}\left(q_{1a}^{T} C\gamma_{\mu} q_{2b}\right) q_{3c}, \label{eq:Bstar}
\end{align}
which interpolate spin-$1/2$ and spin-$3/2$ baryons, and for the meson part $P=\bar{q}_4\, i\gamma_5\, q_5$ and $V_{\mu}=\bar{q}_4\gamma_{\mu} q_5$. Here $a,b,c$ are color indices and $C$ is the charge-conjugation matrix. The S-wave combinations with definite spin are
\begin{equation}
\begin{aligned}
J^{BP} &= B\,P, \qquad J^{VB} = \gamma^{\nu}\gamma_5\, B\, V_{\nu}, \qquad J^{B^*V} = B^{*}_{\mu} V^{\mu},\\
J^{BV}_{\mu} &= B\,V_{\mu}, \qquad J^{B^*P}_{\mu} = B^{*}_{\mu} P,\\
J^{A}_{\mu} &= \gamma^{\nu}\gamma_5\left(B^{*}_{\mu}V_{\nu}-B^{*}_{\nu}V_{\mu}\right),\\
J^{T}_{\mu\nu} &= B^{*}_{\mu}V_{\nu}+B^{*}_{\nu}V_{\mu},
\end{aligned}
\label{eq:forms}
\end{equation}
where the currents in the first line have $J^P=1/2^-$, $J^{BV}_\mu$, $J^{B^*P}_\mu$, and $J^{A}_{\mu}$ have $3/2^-$, and $J^{T}_{\mu\nu}$ has $5/2^-$.

Each system admits two flavor assignments of the baryon and meson factors. For $cc\bar{c}qq$ these are $\Sigma_c^{(*)}\otimes(\bar{c}c)$ and $\Xi_{cc}^{(*)}\otimes(\bar{c}q)$; for $ccc\bar{c}q$ they are $\Xi_{cc}^{(*)}\otimes(\bar{c}c)$ and $\Omega_{ccc}^{*}\otimes(\bar{c}q)$; for $cccc\bar{c}$ only $\Omega_{ccc}^{*}\otimes(\bar{c}c)$ is possible. We work in the isospin limit, $m_u=m_d$ and $\langle\bar uu\rangle=\langle\bar dd\rangle$. The states of $ccc\bar cq$ and $bbb\bar bq$ then form isospin doublets, and the currents of $cc\bar cqq$ and $bb\bar bqq$ with symmetric light flavor describe isospin-one states, which we evaluate for $qq=uu$. The isospin-zero states of these systems, which involve antisymmetric light flavor and, for the $\Sigma$-type assignment, the $\Lambda_Q$ baryon, are not considered. Since the spin-$1/2$ operator~(\ref{eq:B}) vanishes identically for three identical quarks, the $\Omega_{ccc}$ configurations contribute only through $B^*_\mu$. The resulting 14, 11, and 4 currents are listed in Table~\ref{tab:currents}; the bottom systems are described by the same currents with $c\to b$. When identical charm quarks appear in both hadronic factors, the currents are not independent a priori, since a Fierz rearrangement could map one configuration onto the other. An explicit antisymmetrization over all charm fields shows that every current in Table~\ref{tab:currents} is nonvanishing and that the currents within each $J^P$ channel are linearly independent.

\begin{table*}[t]
\caption{Interpolating currents $J_i$. Each row corresponds to one Lorentz structure of Eq.~(\ref{eq:forms}); the entries give the current index and the baryon--meson configuration. The $\Omega_{ccc}$ configurations contain only the spin-$3/2$ baryon. The currents of the bottom systems follow from $c\to b$, i.e., $\Sigma_c\to\Sigma_b$, $\Xi_{cc}\to\Xi_{bb}$, $\Omega_{ccc}\to\Omega_{bbb}$, $\eta_c\to\eta_b$, $J/\psi\to\Upsilon$, and $\bar D^{(*)}\to B^{(*)}$, where $B^{(*)}$ denotes the $\bar bq$ meson.}
\label{tab:currents}
\renewcommand{\arraystretch}{1.2}
\begin{tabular*}{\textwidth}{@{\extracolsep{\fill}}llccc@{}}
\toprule
Structure & $J^P$ & $cc\bar{c}qq$ & $ccc\bar{c}q$ & $cccc\bar{c}$ \\
\midrule
$BP$      & $1/2^-$ & $J_1$: $\Sigma_c\eta_c$, \quad $J_4$: $\Xi_{cc}\bar D$ & $J_1$: $\Xi_{cc}\eta_c$ & -- \\
$VB$      & $1/2^-$ & $J_2$: $\Sigma_cJ/\psi$, \quad $J_5$: $\Xi_{cc}\bar D^*$ & $J_2$: $\Xi_{cc}J/\psi$ & -- \\
$B^*V$    & $1/2^-$ & $J_3$: $\Sigma_c^*J/\psi$, \quad $J_6$: $\Xi_{cc}^*\bar D^*$ & $J_3$: $\Xi_{cc}^*J/\psi$, \quad $J_4$: $\Omega_{ccc}^*\bar D^*$ & $J_1$: $\Omega_{ccc}^*J/\psi$ \\
\midrule
$BV$      & $3/2^-$ & $J_7$: $\Sigma_cJ/\psi$, \quad $J_{10}$: $\Xi_{cc}\bar D^*$ & $J_5$: $\Xi_{cc}J/\psi$ & -- \\
$B^*P$    & $3/2^-$ & $J_8$: $\Sigma_c^*\eta_c$, \quad $J_{11}$: $\Xi_{cc}^*\bar D$ & $J_6$: $\Xi_{cc}^*\eta_c$, \quad $J_8$: $\Omega_{ccc}^*\bar D$ & $J_2$: $\Omega_{ccc}^*\eta_c$ \\
$A$       & $3/2^-$ & $J_9$: $\Sigma_c^*J/\psi$, \quad $J_{12}$: $\Xi_{cc}^*\bar D^*$ & $J_7$: $\Xi_{cc}^*J/\psi$, \quad $J_9$: $\Omega_{ccc}^*\bar D^*$ & $J_3$: $\Omega_{ccc}^*J/\psi$ \\
\midrule
$T$       & $5/2^-$ & $J_{13}$: $\Sigma_c^*J/\psi$, \quad $J_{14}$: $\Xi_{cc}^*\bar D^*$ & $J_{10}$: $\Xi_{cc}^*J/\psi$, \quad $J_{11}$: $\Omega_{ccc}^*\bar D^*$ & $J_4$: $\Omega_{ccc}^*J/\psi$ \\
\bottomrule
\end{tabular*}
\end{table*}

\subsection{Correlation functions and hadronic representation}\label{sec:hadronic}

For two currents of the same spin-parity we consider the correlation function
\begin{equation}
\Pi^{ij}(p)=i\int d^4x\, e^{ip\cdot x}\,\langle 0|\mathcal{T}\big\{J_i(x)\,\bar{J}_j(0)\big\}|0\rangle ,
\label{eq:corr}
\end{equation}
where the Lorentz indices of the $3/2$ and $5/2$ currents are suppressed. The full correlator matrices contain 45, 28, and 5 independent elements for $N_h=3$, 4, and 5, respectively. A current with spin $J$ couples to states of both parities. With $\langle 0|J|P^-(p,s)\rangle=\lambda_-u(p,s)$ and $\langle 0|J|P^+(p,s)\rangle=\lambda_+\gamma_5u(p,s)$, where $u$ is the Dirac, Rarita--Schwinger, or spin-$5/2$ spinor, the pole contributions to the coefficients of the structures $\pslash\,\mathcal{O}_J$ and $\mathcal{O}_J$, with $\mathcal{O}_{1/2}=1$, $\mathcal{O}_{3/2}=g_{\mu\nu}$, and $\mathcal{O}_{5/2}=g_{\mu\alpha}g_{\nu\beta}+g_{\mu\beta}g_{\nu\alpha}$, are
\begin{equation}
\Pi^{\rm had}(p)=\kappa_J\left[\lambda_-^2\,\frac{\pslash+m_-}{m_-^2-p^2}+\lambda_+^2\,\frac{\pslash-m_+}{m_+^2-p^2}\right]\mathcal{O}_J+\ldots ,
\label{eq:had}
\end{equation}
where $\kappa_J=1,-1,1/2$ follows from the spin sums and the ellipsis denotes higher states and the continuum. These structures receive no contributions from states of lower spin. Denoting the spectral densities of the two structures by $\rho_{A}(s)$ and $\rho_{B}(s)$, the combination $\sqrt{s}\,\rho_{A}(s)+\rho_{B}(s)$ does not receive contributions from the positive-parity pole~\cite{Jido:1996ia}, and the Borel-transformed sum rule for the negative-parity state, with $m\equiv m_-$ and $\lambda\equiv\lambda_-$, reads
\begin{equation}
\begin{aligned}
2\kappa_J\,m\,\lambda^2\, e^{-m^2/M^2}&=\int_{s_{\rm th}}^{s_0}ds\, e^{-s/M^2}\left[\sqrt{s}\,\rho_{A}(s)+\rho_{B}(s)\right]\\
&\quad+\Pi_{\rm B}(M^2)\equiv\Pi(M^2,s_0),
\end{aligned}
\label{eq:sumrule}
\end{equation}
where $s_0$ is the continuum threshold and $\Pi_{\rm B}(M^2)$ collects the contributions that cannot be written as a positive spectral density (see below). The mass follows from
\begin{equation}
m^2=-\frac{\partial\ln\Pi(M^2,s_0)}{\partial(1/M^2)} .
\label{eq:mass}
\end{equation}
For a mixed current $J(\beta)=J_i+\beta J_j$ the same relations hold with $\Pi\to\Pi^{ii}+2\beta\,{\rm Re}\,\Pi^{ij}+\beta^2\Pi^{jj}$.

\subsection{Operator product expansion}\label{sec:ope}

On the QCD side the correlation function is expressed through Wick contractions of the quark fields. Contractions of a charm field with an anticharm field at the same point describe annihilation into gluons; they are OZI suppressed and are not included. The number of contraction terms is 284, 312, and 288 for $N_h=3$, 4, and 5. The light- and charm-quark propagators in coordinate space are~\cite{Balitsky:1987bk,Belyaev:1985wza}
\begin{widetext}
\begin{align}
S_q^{ij}(x)&=\frac{i\xslash}{2\pi^2x^4}\delta^{ij}-\frac{m_q}{4\pi^2x^2}\delta^{ij}
-\frac{\langle\bar qq\rangle}{12}\delta^{ij}+\frac{i\xslash m_q\langle\bar qq\rangle}{48}\delta^{ij}
-\frac{x^2\langle\bar q g_s\sigma Gq\rangle}{192}\delta^{ij}+\frac{ix^2\xslash m_q\langle\bar q g_s\sigma Gq\rangle}{1152}\delta^{ij}
-\frac{x^4\langle\bar qq\rangle\langle g_s^2G^2\rangle}{27648}\delta^{ij}\nonumber\\
&\quad-\frac{1}{8}\langle\bar q^j\sigma^{\alpha\beta}q^i\rangle\sigma_{\alpha\beta}
-\frac{ig_sG^{n}_{\alpha\beta}t^n_{ij}}{32\pi^2x^2}\big(\xslash\sigma^{\alpha\beta}+\sigma^{\alpha\beta}\xslash\big)+\cdots,
\label{eq:Sq}\\[4pt]
S_{c}(x)&=\frac{m_{c}^{2}}{4 \pi^{2}} \bigg[ \frac{K_{1}\big(m_{c}\sqrt{-x^{2}}\big) }{\sqrt{-x^{2}}}
+i\frac{\xslash~K_{2}\big( m_{c}\sqrt{-x^{2}}\big)}
{(\sqrt{-x^{2}})^{2}}\bigg]
+\frac{m_{c}\,g_{s} }{32\pi ^{2}}  \int_0^1 dv \,G^{\mu \nu}(vx)\bigg[
    i\,\frac{K_{1}\big( m_{c}\sqrt{-x^{2}}\big) }{\sqrt{-x^{2}}}(\sigma _{\mu \nu }\xslash
+\xslash\sigma _{\mu \nu })\nonumber\\
&\quad+2\sigma_{\mu \nu }K_{0}\big( m_{c}\sqrt{-x^{2}}\big)\bigg]+\cdots,
 \label{eq:Sc}
\end{align}
\end{widetext}
where $K_n$ are modified Bessel functions of the second kind, $t^n=\lambda^n/2$ with the Gell-Mann matrices $\lambda^n$, and the bottom-quark propagator follows from $m_c\to m_b$. The charm-quark propagator is used in the Fock--Schwinger gauge. In addition to the terms shown in Eq.~(\ref{eq:Sc}), we retain the one-gluon term expanded up to two covariant derivatives of the field strength, the two-gluon term, and the gluon-condensate term; the vertices with derivatives and the open two-gluon vertex are retained on heavy lines only. For the light quarks we set $m_q=0$ in Eq.~(\ref{eq:Sq}) and include the free propagator, the one-gluon term, the quark condensate, the mixed condensate through its $x^2$ term, and the soft quark--gluon term; the term proportional to $\langle\bar q^j\sigma^{\alpha\beta}q^i\rangle$ does not contribute to the vacuum correlator.

The condensate content of the expansion depends on the number of light lines. For $cc\bar cqq$ we include the perturbative term and $\langle\bar qq\rangle$, $\langle g_s^2G^2\rangle$, $m_0^2\langle\bar qq\rangle$, $\langle\bar qq\rangle^2$, $\langle g_s^3G^3\rangle$, and $\langle J^2\rangle$, together with the products $\langle\bar qq\rangle\langle g_s^2G^2\rangle$, $m_0^2\langle\bar qq\rangle^2$, $m_0^2\langle\bar qq\rangle\langle g_s^2G^2\rangle$, $(m_0^2\langle\bar qq\rangle)^2$, and $\langle\bar qq\rangle^2\langle g_s^2G^2\rangle$ up to dimension 10. With a single light line the products requiring two light-quark condensates are absent, and the expansion for $ccc\bar cq$ contains dimensions 0, 3, 4, 5, 6, 7, and 9. For $cccc\bar c$ only gluonic operators of dimension 4 and 6 contribute at this order, the latter through $\langle g_s^3G^3\rangle$ and through $\langle J^2\rangle$, which is related to the quark condensate by the equations of motion~\cite{Novikov:1983gd}. The dimension-6 gluonic operators are not multiplied by light-quark condensates, and in the products of dimension 7 to 10 each propagator carries at most one condensate insertion. The four-gluon condensate $\langle g_s^2G^2\rangle^2$ is set to zero; for $cccc\bar c$, where the gluon condensate is most important, its effect is estimated in Sec.~\ref{sec:res5}.

After the Fourier transform, a contribution with $N_h$ heavy lines contains a product of $N_h$ modified Bessel functions, which we represent through $N_h$ Feynman parameters $y_i$ on the simplex $\sum_i y_i=1$. The spectral densities then take the form
\begin{equation}
\rho(s)=\sum_{n}\int_{\Delta_{N_h}}\!d\mu(y)\; f_n(y,s)\,\theta\!\left(s-m_c^2\,\sigma(y)\right),
\label{eq:rho}
\end{equation}
with $\sigma(y)=\sum_{i=1}^{N_h}1/y_i$, polynomial functions $f_n$ and the threshold $s_{\rm th}=N_h^2m_c^2$ at the center of the simplex. Terms whose Borel transform contains a negative power of $M^2$ or a $\delta$-type spectral function are kept in the Borel form $\Pi_{\rm B}(M^2)$ of Eq.~(\ref{eq:sumrule}), and no continuum is subtracted from them. An explicit example is given in Appendix~\ref{app:rho}. The $(N_h-1)$-dimensional integrals are evaluated by a radial decomposition of the simplex around its center, combined with a one-dimensional quadrature in $\sigma$. Near the threshold the integrand behaves as $(\sigma-N_h^2)^{(N_h-3)/2}$, and for $N_h=4$ the square-root behavior is removed by the substitution $\sigma=a+(b-a)u^2$. All steps of the computation were checked independently: the Wick contractions against a dense tensor evaluation, the decomposition into Lorentz structures against a direct numerical evaluation of the correlators at a general space-time point for each condensate, and the Borel transform for $N_h$ heavy lines against iterated two-body convolutions of scalar propagators. The relative deviations are at the level of $10^{-14}$ in the first two cases and below $2\times10^{-7}$ in the last one.

\section{Numerical analysis}\label{sec:numerics}

The input parameters are collected in Table~\ref{tab:inputs}. In the error analysis $\alpha_s(M_Z)$ is varied to $0.1202$, $\rho_c$ to $1/2$~fm, and $\mu$ to $1.5$ and $3$~GeV. The light-quark condensates are evolved to $\mu$ with the same $\alpha_s$, $\langle g_s^3G^3\rangle$ is estimated from the relation $\langle g_s^3G^3\rangle=12\langle g_s^2G^2\rangle/(5\rho_c^2)$~\cite{Schafer:1996wv}, and $\langle J^2\rangle=-\tfrac{4}{9}g_s^4\langle\bar qq\rangle^2\big[2+(\langle\bar ss\rangle/\langle\bar qq\rangle)^2\big]$~\cite{Novikov:1983gd}.

\begin{table}[t]
\caption{Input parameters. The quark condensate is evolved from $1$~GeV to $\mu=2$~GeV. For the bottom quark we use $\barm_b=4.18$~GeV and vary it by $^{+0.04}_{-0.03}$~GeV, a range that contains the value of Ref.~\cite{ParticleDataGroup:2024cfk}.}
\label{tab:inputs}
\renewcommand{\arraystretch}{1.2}\small
\begin{tabular*}{\columnwidth}{@{\extracolsep{\fill}}lll@{}}
\toprule
Parameter & Value & Ref. \\
\midrule
$\barm_c(\barm_c)$ & $1.2730\pm0.0046$~GeV & \cite{ParticleDataGroup:2024cfk} \\
$\barm_b(\barm_b)$ & $4.183\pm0.007$~GeV & \cite{ParticleDataGroup:2024cfk} \\
$\alpha_s(M_Z)$ & $0.1178$ & \cite{ParticleDataGroup:2024cfk} \\
$\langle\bar qq\rangle(1~{\rm GeV})$ & $-(0.255\pm0.008)^3$~GeV$^3$ & \cite{Ioffe:2005ym} \\
$m_0^2$ & $0.8\pm0.2$~GeV$^2$ & \cite{Ioffe:2005ym} \\
$\langle\bar ss\rangle/\langle\bar qq\rangle$ & $0.8\pm0.1$ & \cite{Ioffe:2005ym} \\
$\langle g_s^2G^2\rangle$ & $0.80\pm0.04$~GeV$^4$ & \cite{Narison:2018nbv} \\
$\rho_c$ & $1/3$~fm & \cite{Schafer:1996wv} \\
$\mu$ & $2$~GeV & \\
\bottomrule
\end{tabular*}
\end{table}

The working region in $M^2$ is bounded from above by the requirement that the pole contribution,
\begin{equation}
{\rm PC}(M^2,s_0)=\frac{\Pi(M^2,s_0)}{\Pi(M^2,\infty)},
\label{eq:pc}
\end{equation}
be at least $40\%$, and from below by the convergence of the expansion. For $cc\bar cqq$ and $ccc\bar cq$ we require the highest computed dimension, 10 and 9 respectively, to contribute at most $10\%$ of $\Pi(M^2,s_0)$. In these systems the leading condensates are large by construction, most notably the quark condensate in the chirally odd structure, while the higher terms decrease regularly. For $cccc\bar c$ the highest dimension, 6, is much smaller than dimension 4, so that a bound on dimension 6 alone would not test the convergence; here we require both $|d_4|\leq10\%$ and $|d_6|\leq10\%$. Within the resulting region we demand Borel stability, $|dm/dM^2|\leq0.1$~GeV/GeV$^2$, and quote the mass at the center of the stable part of the window. The continuum threshold is fixed self-consistently by $\sqrt{s_0}=m+0.5$~GeV. The same criteria are used for the bottom systems, whose working regions lie at $M^2\approx8.5$--$16.5$~GeV$^2$. No bound is tuned to open a window; when the self-consistency condition has no solution, the results are given as functions of $s_0$.

\begin{figure*}[t]
\centering
\includegraphics[width=0.98\textwidth]{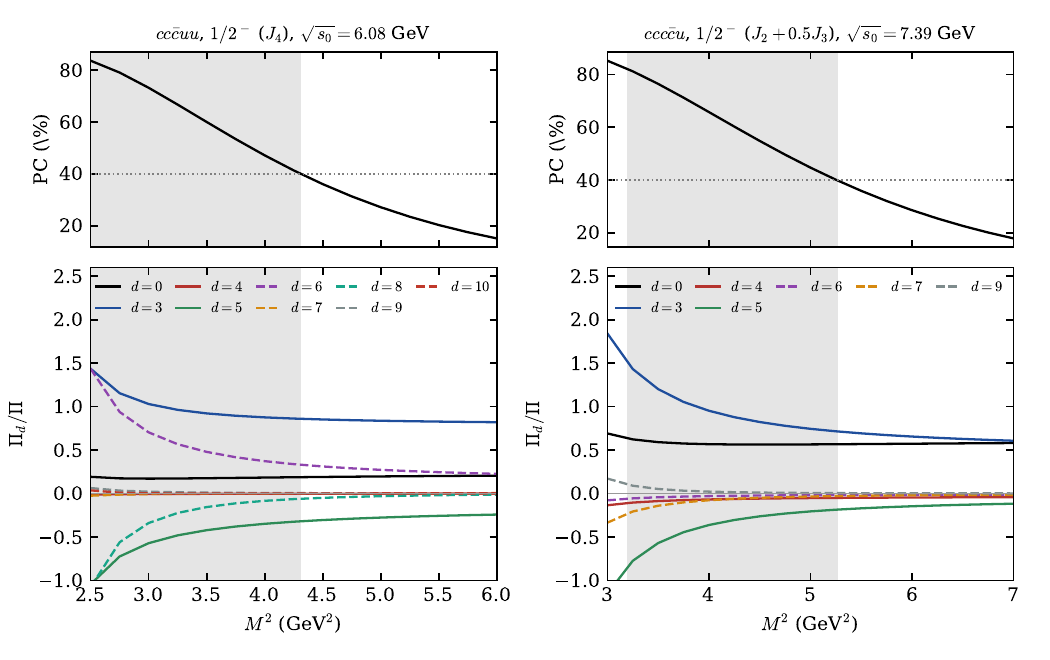}
\caption{Pole contribution (upper panels) and contributions of the individual dimensions relative to $\Pi(M^2,s_0)$ (lower panels) for the $1/2^-$ ground-state channels of $cc\bar cqq$ (left) and $ccc\bar cq$ (right) at their self-consistent thresholds. The shaded bands indicate the working regions.}
\label{fig:ope}
\end{figure*}

The convergence pattern is illustrated in Fig.~\ref{fig:ope} for the ground-state channels of $cc\bar cqq$ and $ccc\bar cq$. As an example, for $ccc\bar cq$ at $s_0=54$~GeV$^2$ and $M^2=5$~GeV$^2$ the shares of dimensions 0, 3, 4, 5, 6, 7, and 9 in the $\Xi_{cc}\eta_c$ correlator are $0.50$, $0.89$, $-0.03$, $-0.31$, $-0.04$, $-0.02$, and $0.007$.

In each channel the mixing parameter of $J(\beta)=J_i+\beta J_j$ is scanned over $\beta\in[-5,5]$ for all pairs of currents at a fixed threshold. A minimum of $m(\beta)$ is accepted if $|\beta_{\min}|<4.95$ and all points within $\beta_{\min}\pm0.5$ have valid working regions; accepted minima are recomputed at $\beta_{\min}$ and $\beta_{\min}\pm0.2$ with the self-consistent threshold.

To determine whether two currents of the same $J^P$ are dominated by the same state we use the overlap ratio
\begin{equation}
R_{ij}(M^2)=\frac{\big[\Pi^{ij}(M^2,s_0)\big]^2}{\Pi^{ii}(M^2,s_0)\,\Pi^{jj}(M^2,s_0)} ,
\label{eq:overlap}
\end{equation}
which approaches unity when both correlators are saturated by a single common pole and is small when the currents couple to different states. If a current $J_a$ overlaps with a current $J_b$ that is dominated by a single lower state, the combination $J_a-cJ_b$ with $c=\Pi^{ba}/\Pi^{bb}$ removes the contribution of this state; this construction is meaningful only if $c$ is independent of $M^2$.

The uncertainties are obtained by varying each input of Table~\ref{tab:inputs} separately at the central threshold and Borel mass, by varying $\sqrt{s_0}$ by $\pm0.1$~GeV and $M^2$ over the stable part of the window, and, for mixed currents, $\beta$ by $\pm0.2$; upward and downward shifts are added in quadrature separately. For splittings within a configuration all variations are applied simultaneously to both states, the $M^2$ variation by evaluating both states at the same relative position within their stable windows. The members of a configuration share the quark content, the condensate structure, and the continuum model, so that the Borel-mass dependence is largely common to them. Treating the $M^2$ variations of the two states as independent yields uncertainties of up to about $40$~MeV for charm and $100$~MeV for bottom, which we regard as a conservative upper bound.

\section{Results and discussion}\label{sec:results}

\subsection{The $cc\bar cqq$ and $ccc\bar cq$ systems}\label{sec:res34}

For $cc\bar cqq$ and $ccc\bar cq$ all diagonal correlators admit self-consistent solutions, and the resulting masses are collected in Table~\ref{tab:diag}, arranged by baryon--meson configuration. The self-consistent thresholds lie at $\sqrt{s_0}=6.08$--$6.65$~GeV and $7.20$--$7.65$~GeV, respectively. The working regions are $0.07$--$2.0$~GeV$^2$ wide for $cc\bar cqq$ and $2.0$--$2.5$~GeV$^2$ wide for $ccc\bar cq$; in the latter system the masses vary by $22$--$45$~MeV over their stable parts, and lowering the stability bound to $0.05$~GeV/GeV$^2$ changes them by at most $3$~MeV. In this system the perturbative and quark-condensate contributions to $\Pi^{13}$ and $\Pi^{34}$ vanish identically, and $\Pi^{89}$ receives contributions only at dimension 9.

\begin{table}[t]
\caption{Masses (GeV) extracted from the diagonal correlators, ordered by baryon--meson configuration, together with the threshold $m_B+m_M$ (GeV) of the configuration; the current index is given in parentheses. Each value is obtained at its self-consistent threshold $\sqrt{s_0}=m+0.5$~GeV. Thresholds marked with $^\dagger$ contain the lattice masses of $\Xi^*_{cc}$ and $\Omega_{ccc}$~\cite{Brown:2014ena}; all other hadron masses are from Ref.~\cite{ParticleDataGroup:2024cfk}.}
\label{tab:diag}
\renewcommand{\arraystretch}{1.2}\setlength{\tabcolsep}{3pt}\small
\begin{tabular*}{\columnwidth}{@{\extracolsep{\fill}}lcccc@{}}
\toprule
Configuration & $m_B+m_M$ & $1/2^-$ & $3/2^-$ & $5/2^-$ \\
\midrule
\multicolumn{5}{@{}l}{$cc\bar cqq$} \\
$\Sigma_c\eta_c$          & 5.438 & 5.680 $(J_1)$ & -- & -- \\
$\Sigma_cJ/\psi$          & 5.551 & 5.712 $(J_2)$ & 5.684 $(J_7)$ & -- \\
$\Sigma_c^*\eta_c$        & 5.503 & -- & 5.818 $(J_8)$ & -- \\
$\Sigma_c^*J/\psi$        & 5.615 & 5.849 $(J_3)$ & 5.878 $(J_9)$ & 5.889 $(J_{13})$ \\
$\Xi_{cc}\bar D$          & 5.486 & 5.579 $(J_4)$ & -- & -- \\
$\Xi_{cc}\bar D^*$        & 5.628 & 5.871 $(J_5)$ & 5.912 $(J_{10})$ & -- \\
$\Xi_{cc}^*\bar D$        & 5.557$^\dagger$ & -- & 5.834 $(J_{11})$ & -- \\
$\Xi_{cc}^*\bar D^*$      & 5.699$^\dagger$ & 6.149 $(J_6)$ & 5.942 $(J_{12})$ & 6.049 $(J_{14})$ \\
\midrule
\multicolumn{5}{@{}l}{$ccc\bar cq$} \\
$\Xi_{cc}\eta_c$          & 6.606 & 6.932 $(J_1)$ & -- & -- \\
$\Xi_{cc}J/\psi$          & 6.719 & 6.905 $(J_2)$ & 6.907 $(J_5)$ & -- \\
$\Xi_{cc}^*\eta_c$        & 6.676$^\dagger$ & -- & 7.085 $(J_6)$ & -- \\
$\Xi_{cc}^*J/\psi$        & 6.789$^\dagger$ & 7.111 $(J_3)$ & 7.017 $(J_7)$ & 7.150 $(J_{10})$ \\
$\Omega_{ccc}^*\bar D$    & 6.661$^\dagger$ & -- & 6.697 $(J_8)$ & -- \\
$\Omega_{ccc}^*\bar D^*$  & 6.803$^\dagger$ & 7.033 $(J_4)$ & 7.030 $(J_9)$ & 7.025 $(J_{11})$ \\
\bottomrule
\end{tabular*}
\end{table}

The lowest states are determined from the mixing analysis. For $cc\bar cqq$ the lowest $1/2^-$ and $5/2^-$ masses are obtained from the pure currents $J_4$ ($\Xi_{cc}\bar D$) and $J_{13}$ ($\Sigma^*_cJ/\psi$), while in the $3/2^-$ channel the combination $J_7-0.4J_{10}$ of the two $BV$ currents lies below all pure currents. For $ccc\bar cq$ the scan accepts the $1/2^-$ minima of $(J_2,J_3)$ at $\beta=0.5$, $(J_2,J_4)$ at $\beta=-0.3$, and $(J_3,J_4)$ at $\beta=-1.5$, and the $3/2^-$ minima of $(J_5,J_9)$ at $\beta=-0.3$ and $(J_7,J_9)$ at $\beta=-0.6$. All pairs containing $J_8$ ($\Omega^*_{ccc}\bar D$) reach their minimum at the pure current $J_8$, which therefore gives the lowest $3/2^-$ state. The apparent $5/2^-$ minimum of $(J_{10},J_{11})$ at $\beta=0.6$ is not accepted, because for $\beta=0.3$--$0.5$ the pole contribution drops below $40\%$ and the working region closes; the minimum lies at the edge of this region. The recomputation of the $(J_2,J_3)$ mixture gives $6.891$~GeV at $\beta=0.5$, and $6.892$ and $6.905$~GeV at $\beta=0.3$ and $0.7$.

\begin{figure*}[t]
\centering
\includegraphics[width=0.98\textwidth]{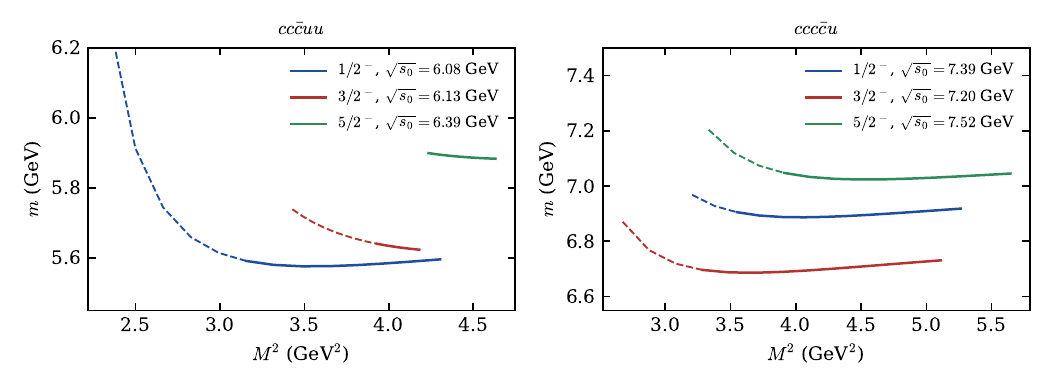}
\caption{Dependence of the ground-state masses on $M^2$ at the self-consistent thresholds for $cc\bar cqq$ (left) and $ccc\bar cq$ (right). Solid lines show the stable parts of the working regions, dashed lines the remaining parts.}
\label{fig:borel}
\end{figure*}

The resulting ground states are summarized in Table~\ref{tab:ground}, their mass differences in Table~\ref{tab:gsplit}, and their Borel-mass dependence is shown in Fig.~\ref{fig:borel}. For $cc\bar cqq$ the uncertainties are dominated by $m_0^2$ ($75$--$160$~MeV) and by the threshold ($45$--$50$~MeV), whereas the ratio $\langle\bar ss\rangle/\langle\bar qq\rangle$ is irrelevant. For $ccc\bar cq$ the largest contributions come from $m_0^2$ ($26$--$87$~MeV) and the threshold (about $45$~MeV), followed by the variation over the stable window ($22$--$33$~MeV); the other inputs contribute about $10$~MeV or less. The dependence on the pole-contribution bound is moderate. For $cc\bar cqq$, lowering the bound to $30\%$ changes the masses by $-20$ to $14$~MeV, and with ${\rm PC}\geq50\%$ only the $1/2^-$ channel retains a self-consistent solution, at $5.574$~GeV. For $ccc\bar cq$ solutions exist for ${\rm PC}\geq50\%$ as well, and between $30\%$ and $50\%$ the masses change by at most $36$~MeV.

\begin{table*}[t]
\caption{Ground states of $cc\bar cqq$ and $ccc\bar cq$: interpolating current, self-consistent threshold, stable part of the Borel window, mass, and residue.}
\label{tab:ground}
\renewcommand{\arraystretch}{1.3}
\begin{tabular*}{\textwidth}{@{\extracolsep{\fill}}llccccc@{}}
\toprule
System & $J^P$ & Current & $\sqrt{s_0}$ (GeV) & $M^2$ (GeV$^2$) & $m$ (GeV) & $\lambda$ ($10^{-3}$~GeV$^6$) \\
\midrule
\multirow{3}{*}{$cc\bar{c}qq$} & $1/2^-$ & $J_4$ & 6.08 & 3.16--4.30 & $5.579^{+0.110}_{-0.094}$ & $4.57^{+0.70}_{-0.60}$ \\
 & $3/2^-$ & $J_7-0.4J_{10}$ & 6.13 & 3.93--4.18 & $5.630^{+0.161}_{-0.143}$ & $3.86^{+0.85}_{-0.60}$ \\
 & $5/2^-$ & $J_{13}$ & 6.39 & 4.24--4.63 & $5.889^{+0.169}_{-0.148}$ & $4.38^{+0.80}_{-0.64}$ \\
\midrule
\multirow{3}{*}{$ccc\bar{c}q$} & $1/2^-$ & $J_2+0.5J_3$ & 7.39 & 3.55--5.26 & $6.891^{+0.078}_{-0.069}$ & $18.9^{+2.6}_{-2.3}$ \\
 & $3/2^-$ & $J_8$ & 7.20 & 3.28--5.11 & $6.697^{+0.063}_{-0.057}$ & $9.4^{+1.3}_{-1.1}$ \\
 & $5/2^-$ & $J_{11}$ & 7.53 & 3.91--5.64 & $7.025^{+0.102}_{-0.086}$ & $19.8^{+2.8}_{-2.4}$ \\
\bottomrule
\end{tabular*}
\end{table*}

\begin{table}[t]
\caption{Mass differences (MeV) between the ground states of Table~\ref{tab:ground}, with correlated input and threshold variations.}
\label{tab:gsplit}
\renewcommand{\arraystretch}{1.35}
\begin{tabular*}{\columnwidth}{@{\extracolsep{\fill}}lccc@{}}
\toprule
System & $m_{3/2}-m_{1/2}$ & $m_{5/2}-m_{3/2}$ & $m_{5/2}-m_{1/2}$ \\
\midrule
$cc\bar cqq$ & $51^{+61}_{-61}$ & $259^{+22}_{-24}$ & $310^{+69}_{-65}$ \\
$ccc\bar cq$ & $-194^{+40}_{-42}$ & $328^{+70}_{-57}$ & $134^{+43}_{-38}$ \\
\bottomrule
\end{tabular*}
\end{table}

For $ccc\bar cq$ the $3/2^-$ ground state lies $194$~MeV below the $1/2^-$ one. This ordering rests on the correlator $\Pi^{88}$, whose condensate structure differs from that of the other channels: within its working region the perturbative share is about $0.39$ and the dimension-5 share lies between $-0.02$ and $-0.22$, compared with $-0.3$ to $-0.7$ in the $\Omega^*_{ccc}\bar D^*$ correlator $\Pi^{99}$. The correlators of $J_8$ and $J_{11}$ contain no quark exchange between the baryon and the meson and are exactly factorizable at this order (Sec.~\ref{sec:conn}), so that the $3/2^-$ and $5/2^-$ ground states of $ccc\bar cq$ cannot be distinguished from the $\Omega^*_{ccc}\bar D^{(*)}$ two-hadron states. Since the three ground states of each system originate from different baryon--meson configurations, the mass differences in Table~\ref{tab:gsplit} do not represent splittings within a spin multiplet; these are discussed in Sec.~\ref{sec:multiplets}.

\subsection{The $cccc\bar c$ system}\label{sec:res5}

In the fully charmed system the gluon condensate gives a large contribution at small $M^2$. Without continuum subtraction its ratio to the perturbative term in the $\Omega^*_{ccc}\eta_c$ correlator is $-0.19$ at $M^2=4$~GeV$^2$ and $-0.06$ at $M^2=6$~GeV$^2$, and it grows in magnitude when $s_0$ approaches the five-quark threshold. The condition $|d_4|\leq10\%$ thus requires large Borel masses, and the pole condition then requires large thresholds. The gluon-condensate contributions were tested independently through the factorizable Wick terms of $\Pi^{22}$, which reproduce the product of the separately computed $\Omega^*_{ccc}$ and $\eta_c$ correlators to better than $10^{-14}$ for every condensate.

\begin{figure*}[t]
\centering
\includegraphics[width=0.98\textwidth]{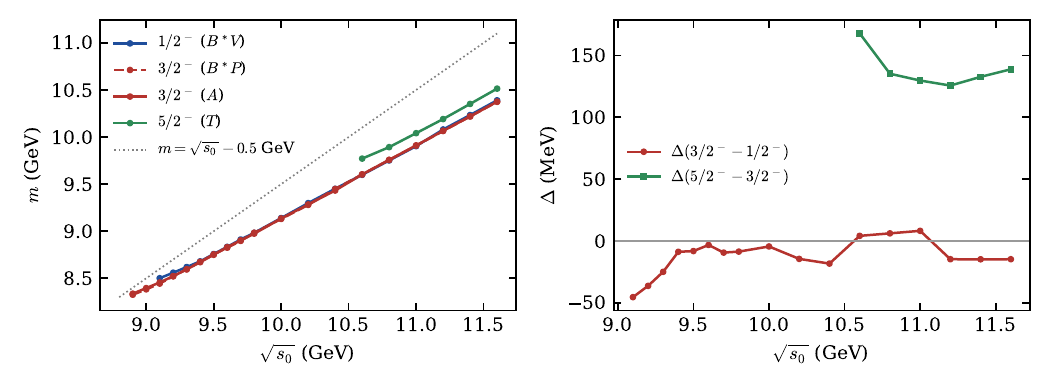}
\caption{$cccc\bar c$: masses obtained from the four currents as functions of $\sqrt{s_0}$ (left) and the splittings between the members of the $\Omega^*_{ccc}J/\psi$ configuration (right). The dotted line indicates the self-consistency condition $m=\sqrt{s_0}-0.5$~GeV.}
\label{fig:nh5}
\end{figure*}

The working regions open at $\sqrt{s_0}\approx9.1$ and $8.9$~GeV in the $1/2^-$ and $3/2^-$ channels and only at $\sqrt{s_0}\approx10.5$~GeV in the $5/2^-$ channel, where the dimension-4 share remains close to $-20\%$ up to $M^2\approx12$~GeV$^2$ at lower thresholds. Wherever a window exists, $m+0.5~{\rm GeV}-\sqrt{s_0}$ is negative, between $-0.06$ and $-0.73$~GeV, so that the self-consistency condition has no solution. The masses grow almost linearly with the threshold, with slopes of $0.74$--$0.79$ between $\sqrt{s_0}=10.6$ and $11.6$~GeV, and no plateau develops (Fig.~\ref{fig:nh5} and Table~\ref{tab:nh5}). We therefore regard the absolute masses as determined by the choice of $s_0$. The two $3/2^-$ currents give masses within $17$~MeV of each other over the full range and within $3$~MeV for $\sqrt{s_0}\geq10.6$~GeV, and their mixing changes the mass by about $1$~MeV for $\beta\in[-3,0]$.

\begin{table}[t]
\caption{$cccc\bar c$: masses obtained from the four currents and splittings between the members of the $\Omega^*_{ccc}J/\psi$ configuration at fixed thresholds; $\Delta_{3/2-1/2}=m(J_3)-m(J_1)$ and $\Delta_{5/2-3/2}=m(J_4)-m(J_3)$. A dash indicates the absence of a working region.}
\label{tab:nh5}
\renewcommand{\arraystretch}{1.15}\setlength{\tabcolsep}{2pt}\small
\begin{tabular*}{\columnwidth}{@{\extracolsep{\fill}}ccccccc@{}}
\toprule
 & \multicolumn{4}{c}{$m$ (GeV)} & \multicolumn{2}{c}{$\Delta$ (MeV)} \\
\cmidrule(lr){2-5}\cmidrule(l){6-7}
$\sqrt{s_0}$ (GeV) & $J_1$ & $J_2$ & $J_3$ & $J_4$ & $\Delta_{3/2-1/2}$ & $\Delta_{5/2-3/2}$ \\
\midrule
8.9  & --     & 8.326  & 8.337  & --     & --    & --  \\
9.1  & 8.501  & 8.444  & 8.456  & --     & $-45$ & --  \\
9.4  & 8.680  & 8.675  & 8.671  & --     & $-9$  & --  \\
9.8  & 8.983  & 8.984  & 8.974  & --     & $-9$  & --  \\
10.2 & 9.299  & 9.280  & 9.284  & --     & $-15$ & --  \\
10.6 & 9.599  & 9.603  & 9.603  & 9.771  & 4     & 168 \\
10.8 & 9.751  & 9.757  & 9.757  & 9.893  & 6     & 135 \\
11.0 & 9.903  & 9.908  & 9.911  & 10.041 & 8     & 130 \\
11.2 & 10.080 & 10.065 & 10.065 & 10.191 & $-15$ & 126 \\
11.4 & 10.234 & 10.218 & 10.219 & 10.352 & $-15$ & 133 \\
11.6 & 10.390 & 10.374 & 10.375 & 10.514 & $-15$ & 139 \\
\bottomrule
\end{tabular*}
\end{table}

The splittings are far more stable than the masses. For $\sqrt{s_0}=10.8$--$11.6$~GeV, where all channels possess working regions, $\Delta(3/2^--1/2^-)$ lies between $-15$ and $8$~MeV and $\Delta(5/2^--3/2^-)$ between $126$ and $139$~MeV; at $\sqrt{s_0}=10.6$~GeV, where the $5/2^-$ window has just opened, the latter amounts to $168$~MeV. The input parameters change $\Delta(5/2^--3/2^-)$ by $^{+8}_{-1}$~MeV and $\Delta(3/2^--1/2^-)$ by less than $1$~MeV. 
The four-gluon condensate, estimated in the factorization approximation with the same rules for the products as in Sec.~\ref{sec:ope}, contributes less than $0.1\%$ in the working regions, shifts the masses by at most $1.6$~MeV, and changes $\Delta(5/2^--3/2^-)$ at $\sqrt{s_0}=11.0$~GeV by $-1.2$~MeV; the working regions remain unchanged. The high thresholds are therefore not an artifact of the truncation at dimension 6.

The $5/2^-$ channel has a further property that is specific to four identical charm quarks. If all quarks occupy a common S-wave orbital, the four charm quarks must be antisymmetric in color and spin. Since they combine with the antiquark into a color singlet, the $cccc$ subsystem forms a color triplet, whose permutation symmetry allows only total spin $S_{cccc}=1$; the total spin $5/2$ is thus excluded. Accordingly, compact quark-model calculations find only $1/2^-$ and $3/2^-$ S-wave states for $cccc\bar c$~\cite{An:2020jix,Zhang:2023hmg,Liang:2024met}. The same restriction can be verified directly for the currents. Projecting the antisymmetrized currents onto the large Dirac components of all five fields, which corresponds to the nonrelativistic S-wave configuration, and diagonalizing the total-spin Casimir operator in the resulting space, one finds $S=1/2$ for $J_1$, $S=3/2$ for $J_2$, and $S=1/2$ and $3/2$ for $J_3$ and $J_4$. The spin-$5/2$ component of $J_4$, which is the one selected by the Lorentz structure of the sum rule, therefore stems exclusively from the small components, i.e., from configurations with internal relative motion. This is consistent with the high threshold required in the $5/2^-$ channel and with the large $5/2^-$--$3/2^-$ splitting. For $cc\bar cqq$ and $ccc\bar cq$ at most three charm quarks are identical, and the S-wave spin-$5/2$ configuration is allowed.

\subsection{The bottom systems}\label{sec:resb}

The bottom systems are described by the same currents and the same operator product expansion, with $\barm_c$ replaced by $\barm_b$. The gluon condensate is much less important than for charm: at $M^2=10$~GeV$^2$ it contributes $1$--$6\%$ of the correlators for $bb\bar bqq$ and $bbb\bar bq$, and for $bbbb\bar b$ its share decreases from $15\%$ at $M^2=10$~GeV$^2$ to below $10\%$ for $M^2\gtrsim12$~GeV$^2$. The resulting masses are collected in Table~\ref{tab:diagb}. Their uncertainties amount to $0.11$--$0.25$~GeV for $bb\bar bqq$ and $bbb\bar bq$, dominated by $m_0^2$ and $\barm_b$, and to about $0.10$~GeV for $bbbb\bar b$, dominated by $\barm_b$ and $s_0$. Since the comparison with the charm systems in Sec.~\ref{sec:multiplets} involves only the pure currents, we do not perform the mixing analysis for the bottom systems and do not assign ground states.

\begin{table}[t]
\caption{Masses (GeV) of the bottom systems from the diagonal correlators at their self-consistent thresholds $\sqrt{s_0}=m+0.5$~GeV, ordered by baryon--meson configuration, together with the threshold $m_B+m_M$ (GeV) of the configuration; the current index is given in parentheses. The thresholds contain the lattice masses of $\Xi_{bb}$, $\Xi^*_{bb}$, and $\Omega_{bbb}$~\cite{Brown:2014ena} and the meson masses of Ref.~\cite{ParticleDataGroup:2024cfk}. For the $\Sigma_b$ configurations and for the $5/2^-$ state of $bbbb\bar b$ no self-consistent solution exists (see text).}
\label{tab:diagb}
\renewcommand{\arraystretch}{1.2}\setlength{\tabcolsep}{3pt}\small
\begin{tabular*}{\columnwidth}{@{\extracolsep{\fill}}lcccc@{}}
\toprule
Configuration & $m_B+m_M$ & $1/2^-$ & $3/2^-$ & $5/2^-$ \\
\midrule
\multicolumn{5}{@{}l}{$bb\bar bqq$} \\
$\Xi_{bb}B$          & 15.422 & 14.587 $(J_4)$ & -- & -- \\
$\Xi_{bb}B^*$        & 15.468 & 14.629 $(J_5)$ & 14.700 $(J_{10})$ & -- \\
$\Xi_{bb}^*B$        & 15.457 & -- & 14.660 $(J_{11})$ & -- \\
$\Xi_{bb}^*B^*$      & 15.503 & 14.775 $(J_6)$ & 14.636 $(J_{12})$ & 14.726 $(J_{14})$ \\
\midrule
\multicolumn{5}{@{}l}{$bbb\bar bq$} \\
$\Xi_{bb}\eta_b$          & 19.542 & 18.497 $(J_1)$ & -- & -- \\
$\Xi_{bb}\Upsilon$        & 19.603 & 18.479 $(J_2)$ & 18.483 $(J_5)$ & -- \\
$\Xi_{bb}^*\eta_b$        & 19.577 & -- & 18.554 $(J_6)$ & -- \\
$\Xi_{bb}^*\Upsilon$      & 19.638 & 18.550 $(J_3)$ & 18.526 $(J_7)$ & 18.565 $(J_{10})$ \\
$\Omega_{bbb}^*B$         & 19.645 & -- & 18.401 $(J_8)$ & -- \\
$\Omega_{bbb}^*B^*$       & 19.691 & 18.534 $(J_4)$ & 18.533 $(J_9)$ & 18.531 $(J_{11})$ \\
\midrule
\multicolumn{5}{@{}l}{$bbbb\bar b$} \\
$\Omega_{bbb}^*\eta_b$    & 23.765 & -- & 22.328 $(J_2)$ & -- \\
$\Omega_{bbb}^*\Upsilon$  & 23.826 & 22.351 $(J_1)$ & 22.334 $(J_3)$ & -- \\
\bottomrule
\end{tabular*}
\end{table}

For $bbb\bar bq$ all eleven currents admit self-consistent solutions, at $\sqrt{s_0}=18.90$--$19.07$~GeV, with working regions of $5.5$--$5.9$~GeV$^2$. For $bb\bar bqq$ this holds for the seven currents of the $\Xi_{bb}$ configurations, at $\sqrt{s_0}=15.09$--$15.28$~GeV, whereas none of the seven currents of the $\Sigma_b$ configurations admits a self-consistent solution. For the latter, the working regions open between $\sqrt{s_0}=15.0$ and $15.2$~GeV. At $\sqrt{s_0}=15.2$~GeV the extracted masses of $14.58$--$14.61$~GeV lie about $0.1$~GeV below $\sqrt{s_0}-0.5$~GeV, and the masses then grow with the threshold with slopes of $0.53$--$0.64$. For these configurations we quote only splittings at a common threshold. In contrast to $cccc\bar c$, the $1/2^-$ and $3/2^-$ channels of $bbbb\bar b$ admit self-consistent solutions, at $\sqrt{s_0}=22.83$--$22.85$~GeV. The $5/2^-$ channel, however, acquires a working region only between $\sqrt{s_0}=23.6$ and $24.0$~GeV, about $1.2$~GeV above the thresholds of the other two channels. This mirrors the behavior found for $cccc\bar c$ in Sec.~\ref{sec:res5}, where the $5/2^-$ window opens about $1.5$~GeV above those of the $1/2^-$ and $3/2^-$ channels.

\subsection{Finite width}\label{sec:width}

The hadronic side of Eq.~(\ref{eq:sumrule}) treats the lowest state as a pole of zero width. Since the charm states lie above their baryon--meson thresholds, we estimate the effect of a finite width by replacing the pole by a Breit--Wigner distribution with an energy-dependent width,
\begin{equation}
\rho_H(s)=\frac{\lambda^2}{\pi}\,\frac{\sqrt{s}\,\Gamma(s)}{(m^2-s)^2+s\,\Gamma^2(s)},\qquad
\Gamma(s)=\Gamma_0\,\frac{m}{\sqrt{s}}\left[\frac{q(s)}{q(m^2)}\right]^{2L+1},
\label{eq:bw}
\end{equation}
so that the left-hand side of Eq.~(\ref{eq:sumrule}) becomes $2\kappa_J\int_{s_{\rm th}}^{s_0}ds\,\sqrt{s}\,\rho_H(s)\,e^{-s/M^2}$. Here $q(s)$ is the momentum in the baryon--meson channel of the current, with threshold $s_{\rm th}=(m_B+m_M)^2$, and $L=0$. The hadron masses are those used for the thresholds in Table~\ref{tab:diag}; for a mixed current the channel of the dominant component is used. For each current we keep $s_0$ and the stable part of the Borel window of the zero-width analysis and determine $m$ by requiring that the effective mass of the hadronic side, defined as in Eq.~(\ref{eq:mass}), reproduce that of the QCD side over this window in the least-squares sense. The reference value for $\Gamma_0\to0$ is obtained in the same way; it agrees with the masses of Tables~\ref{tab:diag} and \ref{tab:ground} within $10$~MeV. We consider $\Gamma_0=10$, $50$, $100$, and $200$~MeV.

\begin{table}[t]
\caption{Shifts $\Delta m=m(\Gamma_0)-m(0)$ (MeV) of the charm ground states of Table~\ref{tab:ground} for the Breit--Wigner distribution of Eq.~(\ref{eq:bw}), at the continuum thresholds and in the stable Borel windows of the zero-width analysis. The third column gives the channel entering $\Gamma(s)$ and its threshold (GeV).}
\label{tab:width}
\renewcommand{\arraystretch}{1.2}\setlength{\tabcolsep}{2pt}\small
\begin{tabular*}{\columnwidth}{@{\extracolsep{\fill}}llccccc@{}}
\toprule
 & & & \multicolumn{4}{c}{$\Gamma_0$ (MeV)} \\
\cmidrule(l){4-7}
System & $J^P$ & Channel & 10 & 50 & 100 & 200 \\
\midrule
\multirow{3}{*}{$cc\bar cqq$} & $1/2^-$ & $\Xi_{cc}\bar D$ (5.486) & $-3$ & $-18$ & $-38$ & $-69$ \\
 & $3/2^-$ & $\Sigma_cJ/\psi$ (5.551) & $-4$ & $-24$ & $-53$ & $-50$ \\
 & $5/2^-$ & $\Sigma^*_cJ/\psi$ (5.615) & 0 & $-2$ & $-2$ & 0 \\
\midrule
\multirow{3}{*}{$ccc\bar cq$} & $1/2^-$ & $\Xi_{cc}J/\psi$ (6.719) & $-1$ & $-6$ & $-10$ & $-17$ \\
 & $3/2^-$ & $\Omega^*_{ccc}\bar D$ (6.661) & $-7$ & $-24$ & $-19$ & $-17$ \\
 & $5/2^-$ & $\Omega^*_{ccc}\bar D^*$ (6.803) & $-1$ & $-3$ & $-4$ & $-5$ \\
\bottomrule
\end{tabular*}
\end{table}

The shifts of the ground states are given in Table~\ref{tab:width}. Over all currents of $cc\bar cqq$ and $ccc\bar cq$ the masses change by at most $7$, $24$, $53$, and $69$~MeV for $\Gamma_0=10$, $50$, $100$, and $200$~MeV, which is within the uncertainties of Table~\ref{tab:ground}. The largest shifts occur for the states closest to their thresholds, the $1/2^-$ and $3/2^-$ ground states of $cc\bar cqq$, which lie $93$ and $79$~MeV above the $\Xi_{cc}\bar D$ and $\Sigma_cJ/\psi$ thresholds, and the $3/2^-$ ground state of $ccc\bar cq$, $36$~MeV above the $\Omega^*_{ccc}\bar D$ threshold. For the two $3/2^-$ states the shift does not grow further for the largest widths. For the $5/2^-$ ground states, which lie $0.22$--$0.27$~GeV above their thresholds, the shifts remain below $5$~MeV. Since the members of a configuration share the channel, the splittings of Table~\ref{tab:mult} change by at most $2$, $10$, $20$, and $39$~MeV for the four values of $\Gamma_0$, the largest change occurring for $\Delta_{3/2-1/2}$ of $\Xi^*_{cc}\bar D^*$. In the bottom systems the position of the states relative to the thresholds depends on the heavy-quark mass scheme (Sec.~\ref{sec:comparison}): with $\barm_b(\barm_b)$ all masses lie below the thresholds of the corresponding channels, where Eq.~(\ref{eq:bw}) does not apply, and with the pole mass above them. Since for $cccc\bar c$ no self-consistent solution exists, the test is restricted to the charm systems with $N_h=3$ and 4. This test assumes that the spectral density below $s_0$ is dominated by a single state and examines how the extracted mass depends on its width. The decomposition of Sec.~\ref{sec:conn} shows that the correlators are dominated by factorizable two-hadron contributions; the weak dependence on $\Gamma_0$ therefore reflects the Borel weight, which makes the extracted mass insensitive to the distribution of the strength below $s_0$, and does not by itself establish a resonance.

\subsection{Spin multiplets}\label{sec:multiplets}

We now compare states of different $J^P$ that are obtained from the same baryon--meson configuration, i.e., from currents that differ only in the Lorentz coupling of Eq.~(\ref{eq:forms}). A $B^*\otimes V$ configuration provides a triplet of $B^*V$, $A$, and $T$ currents, and a $B\otimes V$ configuration a doublet of $VB$ and $BV$ currents. A meaningful comparison requires that the currents of the two configurations of a given system couple to different states, which we examine with the overlap ratios of Eq.~(\ref{eq:overlap}).

\begin{table}[t]
\caption{Overlap ratios $R_{ij}$ of Eq.~(\ref{eq:overlap}) for $M^2=3.5$--$5$~GeV$^2$, with $s_0=41$~GeV$^2$ for $cc\bar cqq$ and $56$~GeV$^2$ for $ccc\bar cq$. The upper part compares the currents with the same Lorentz structure in the two configurations; the lower part compares the axial-type current of the $\Xi^*_{cc}$ configuration with the $BV$ currents.}
\label{tab:overlap}
\renewcommand{\arraystretch}{1.2}\small
\begin{tabular*}{\columnwidth}{@{\extracolsep{\fill}}lcc@{}}
\toprule
 & $cc\bar cqq$ & $ccc\bar cq$ \\
\midrule
 & $\Sigma_c^*J/\psi$ vs $\Xi_{cc}^*\bar D^*$ & $\Xi_{cc}^*J/\psi$ vs $\Omega_{ccc}^*\bar D^*$ \\
\cmidrule(lr){2-2}\cmidrule(l){3-3}
$B^*V$ ($1/2^-$) & 0.001--0.007 & 0.006--0.025 \\
$A$ ($3/2^-$)    & 0.04--0.06   & 0.11--0.15 \\
$T$ ($5/2^-$)    & 0.82--0.97   & 0.95 \\
\midrule
 & \multicolumn{2}{c}{$A$ ($\Xi^*_{cc}$ configuration) vs $BV$} \\
\cmidrule(l){2-3}
$\Xi_{cc}\bar D^*$ / $\Xi_{cc}J/\psi$ & 0.06--0.10 & 0.38--0.41 \\
$\Sigma_cJ/\psi$                      & 0.27--0.31 & -- \\
\bottomrule
\end{tabular*}
\end{table}

The ratios in Table~\ref{tab:overlap} show that in the $1/2^-$ and $3/2^-$ channels the triplet currents of the two configurations are nearly orthogonal, $R\lesssim0.15$. In the $5/2^-$ channel, by contrast, the two tensor currents have $R=0.82$--$0.97$ for $cc\bar cqq$ and $R\approx0.95$ for $ccc\bar cq$, so that they couple predominantly to the same state. In these systems the $5/2^-$ member is shared by the two configurations, and the difference of $125$--$160$~MeV between the masses extracted from the two tensor currents at their self-consistent thresholds, or $62$--$90$~MeV at a common threshold, reflects the dependence of the sum rule on the choice of the current rather than a physical splitting. We include half of this difference at a common threshold, $39$--$45$~MeV for $cc\bar cqq$ and $31$--$41$~MeV for $ccc\bar cq$, as an additional uncertainty of $\Delta(5/2^--3/2^-)$. For $cccc\bar c$ there is only one tensor current, and this ambiguity does not arise.

The axial-type current of the $\Xi$-type triplet overlaps moderately with a $BV$ current, with $R\approx0.4$ for the $\Xi_{cc}J/\psi$ current of the same configuration in $ccc\bar cq$ and $R\approx0.3$ for the $\Sigma_cJ/\psi$ current in $cc\bar cqq$. The $3/2^-$ member of the triplet can thus be pulled toward a lower $3/2^-$ state. For $ccc\bar cq$ the coefficient $c=\Pi^{57}/\Pi^{55}$ is constant within $1.6\%$ for $M^2=3$--$5.5$~GeV$^2$, and the combination $J_7-cJ_5$ gives a self-consistent mass of $7.069$~GeV instead of $7.017$~GeV. For $cc\bar cqq$ the corresponding coefficient varies between $-0.30$ and $-0.40$ over the same range, the $BV$ correlator is not dominated by a single state, and the combination has no working region; there the original current is used.

\begin{table*}[t]
\caption{Charm systems: splittings (MeV) within baryon--meson configurations, $\Delta_{3/2-1/2}=m(3/2^-)-m(1/2^-)$ and $\Delta_{5/2-3/2}=m(5/2^-)-m(3/2^-)$. ``Self-consistent'': each state at its own threshold, with correlated uncertainties (Sec.~\ref{sec:numerics}). ``Common $s_0$'': interval of values obtained over the indicated range of thresholds. For $N_h=3$ and $4$ the values of $\Delta_{5/2-3/2}$ carry an additional uncertainty of $39$--$45$ and $31$--$41$~MeV, respectively, from the $5/2^-$ state shared by the two configurations (Sec.~\ref{sec:multiplets}).}
\label{tab:mult}
\renewcommand{\arraystretch}{1.3}
\begin{tabular*}{\textwidth}{@{\extracolsep{\fill}}lcccccc@{}}
\toprule
 & & \multicolumn{2}{c}{Self-consistent $s_0$} & \multicolumn{3}{c}{Common $s_0$} \\
\cmidrule(lr){3-4}\cmidrule(l){5-7}
Configuration & $N_h$ & $\Delta_{3/2-1/2}$ & $\Delta_{5/2-3/2}$ & $\sqrt{s_0}$ (GeV) & $\Delta_{3/2-1/2}$ & $\Delta_{5/2-3/2}$ \\
\midrule
\multicolumn{7}{@{}l}{\emph{$J/\psi$ chain}} \\
$\Sigma^*_cJ/\psi$ & 3 & $29^{+28}_{-16}$ & $10^{+3}_{-7}$ & 6.3--6.6 & $[6,\,18]$ & $[5,\,6]$ \\
$\Xi^*_{cc}J/\psi$ & 4 & $-94^{+7}_{-12}$ & $133^{+33}_{-18}$ & 7.3--7.7 & $[-61,\,-43]$ & $[64,\,89]$ \\
$\Xi^*_{cc}J/\psi$, with $J_7-cJ_5$ & 4 & $-42^{+3}_{-5}$ & $81^{+27}_{-13}$ & & & \\
$\Omega^*_{ccc}J/\psi$ & 5 & -- & -- & 10.8--11.6 & $[-15,\,8]$ & $[126,\,139]$ \\
\midrule
\multicolumn{7}{@{}l}{\emph{$\bar D^*$ chain}} \\
$\Xi^*_{cc}\bar D^*$ & 3 & $-208^{+7}_{-31}$ & $107^{+29}_{-5}$ & 6.4--6.6 & $[-118,\,-101]$ & $[47,\,59]$ \\
$\Omega^*_{ccc}\bar D^*$ & 4 & $-4^{+5}_{-2}$ & $-5^{+1}_{-1}$ & 7.3--7.7 & $[-3,\,-1]$ & $[-3,\,-2]$ \\
\midrule
\multicolumn{7}{@{}l}{\emph{Doublets}} \\
$\Sigma_cJ/\psi$ & 3 & $-28^{+1}_{-1}$ & -- & 6.2--6.6 & $[-16,\,-13]$ & -- \\
$\Xi_{cc}\bar D^*$ & 3 & $41^{+12}_{-3}$ & -- & 6.2--6.6 & $[19,\,25]$ & -- \\
$\Xi_{cc}J/\psi$ & 4 & $2^{+7}_{-4}$ & -- & 7.3--7.7 & $[1,\,3]$ & -- \\
\bottomrule
\end{tabular*}
\end{table*}

\begin{figure*}[t]
\centering
\includegraphics[width=0.98\textwidth]{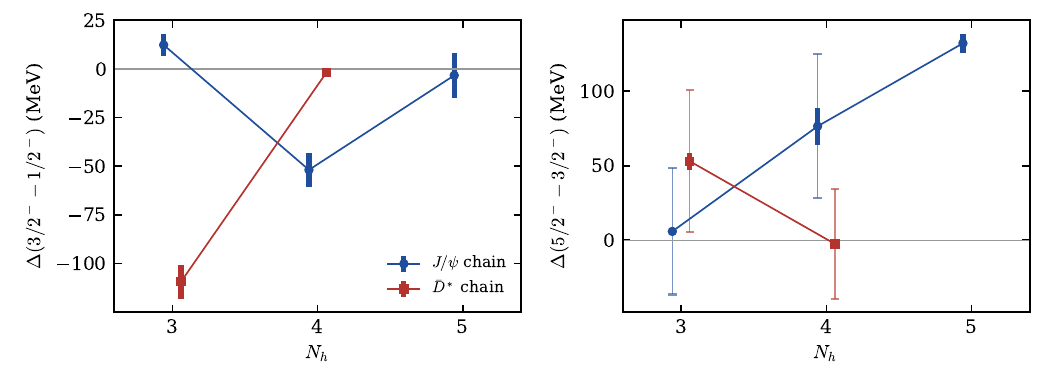}
\caption{Charm systems: splittings within baryon--meson configurations at a common threshold as functions of $N_h$ for the $J/\psi$ chain $\Sigma^*_cJ/\psi\to\Xi^*_{cc}J/\psi\to\Omega^*_{ccc}J/\psi$ and the $\bar D^*$ chain $\Xi^*_{cc}\bar D^*\to\Omega^*_{ccc}\bar D^*$. Thick bars show the ranges over the threshold intervals given in Table~\ref{tab:mult}; thin bars include the uncertainty from the shared $5/2^-$ state.}
\label{fig:split}
\end{figure*}

The splittings within the individual configurations of the charm systems are listed in Table~\ref{tab:mult} and displayed in Fig.~\ref{fig:split}. They are far less sensitive to the input parameters than the splittings between ground states, because $m_0^2$ and $s_0$ shift the members of a configuration in the same direction; changing the pole-contribution bound from $30\%$ to $50\%$ alters them by at most $3$~MeV for $ccc\bar cq$. At a common threshold the splittings are generally smaller than at the self-consistent thresholds, since the latter differ by up to $0.2$~GeV within a multiplet and the masses grow with $s_0$.

In the charm systems two patterns emerge. The $\Sigma^*_cJ/\psi$ and $\Omega^*_{ccc}\bar D^*$ triplets and the three doublets are nearly degenerate, with splittings between about $-30$ and $40$~MeV at the self-consistent thresholds and between $-16$ and $25$~MeV at a common threshold. The triplets containing a $\Xi^*_{cc}$ baryon, $\Xi^*_{cc}\bar D^*$ for $N_h=3$ and $\Xi^*_{cc}J/\psi$ for $N_h=4$, show a non-monotonic pattern with the $3/2^-$ member lowest. For $N_h=4$ about half of the depression of the $3/2^-$ member is removed when the axial-type current is orthogonalized to the doublet state, and a splitting of $-42$~MeV remains. Along the chain~(\ref{eq:chain}) the $5/2^-$--$3/2^-$ splitting of the $J/\psi$ configurations grows from a value compatible with zero at $N_h=3$ to about $75\pm35$~MeV at $N_h=4$ and about $130$~MeV at $N_h=5$, while that of the $\bar D^*$ configurations decreases from about $50$~MeV to zero. A uniform suppression of spin-dependent effects with increasing $N_h$ is therefore not observed, and the dependence on $N_h$ alone does not separate the different sources of the splittings.

\begin{table*}[t]
\caption{Splittings (MeV) within baryon--meson configurations for charm and bottom, $\Delta_{3/2-1/2}=m(3/2^-)-m(1/2^-)$ and $\Delta_{5/2-3/2}=m(5/2^-)-m(3/2^-)$. In all configurations the meson is the vector state. Unless marked otherwise, the values refer to the self-consistent thresholds of the individual states, with correlated uncertainties. Values marked with $^\dagger$ are ranges at a common threshold for configurations without self-consistent solutions: $\sqrt{s_0}=15.2$--$16.0$~GeV for $\Sigma^{(*)}_b(b\bar b)$, $10.8$--$11.6$~GeV for $\Omega^*_{ccc}(c\bar c)$, and $24.0$--$24.8$~GeV for the $5/2^-$ state of $\Omega^*_{bbb}(b\bar b)$. For $N_h=3$ and $4$ the values of $\Delta_{5/2-3/2}$ carry an additional uncertainty from the $5/2^-$ state shared by the two configurations (Sec.~\ref{sec:multiplets}). $^\ddagger$No quark exchange between the baryon and the meson; the correlators are exactly factorizable at this order (Sec.~\ref{sec:conn}).}
\label{tab:cb}
\renewcommand{\arraystretch}{1.35}
\begin{tabular*}{\textwidth}{@{\extracolsep{\fill}}llccccc@{}}
\toprule
 & & & \multicolumn{2}{c}{$Q=c$} & \multicolumn{2}{c}{$Q=b$} \\
\cmidrule(lr){4-5}\cmidrule(l){6-7}
Configuration & Light degrees of freedom & $N_h$ & $\Delta_{3/2-1/2}$ & $\Delta_{5/2-3/2}$ & $\Delta_{3/2-1/2}$ & $\Delta_{5/2-3/2}$ \\
\midrule
\multicolumn{7}{@{}l}{\emph{Light degrees of freedom with a unique spin}} \\
$\Sigma^*_Q(Q\bar Q)$ & $qq$ diquark, $s_{qq}=1$ & 3 & $29^{+28}_{-16}$ & $10^{+3}_{-7}$ & $[-5,\,-3]^\dagger$ & $[1,\,4]^\dagger$ \\
$\Sigma_Q(Q\bar Q)$ & $qq$ diquark, $s_{qq}=1$ & 3 & $-28^{+1}_{-1}$ & -- & $[-5,\,-4]^\dagger$ & -- \\
$\Xi^*_{QQ}(Q\bar Q)$ & $q$ in the baryon & 4 & $-94^{+7}_{-12}$ & $133^{+33}_{-18}$ & $-24^{+5}_{-3}$ & $39^{+11}_{-6}$ \\
$\Xi_{QQ}(Q\bar Q)$ & $q$ in the baryon & 4 & $2^{+7}_{-4}$ & -- & $4^{+13}_{-3}$ & -- \\
$\Omega^*_{QQQ}(\bar Qq)^\ddagger$ & $q$ in the meson & 4 & $-4^{+5}_{-2}$ & $-5^{+1}_{-1}$ & $-2^{+2}_{-0}$ & $-1^{+2}_{-0}$ \\
\midrule
\multicolumn{7}{@{}l}{\emph{Two light quarks in different hadrons}} \\
$\Xi^*_{QQ}(\bar Qq)$ & $q$ in the baryon and in the meson & 3 & $-208^{+7}_{-31}$ & $107^{+29}_{-5}$ & $-139^{+9}_{-62}$ & $91^{+72}_{-10}$ \\
$\Xi_{QQ}(\bar Qq)$ & $q$ in the baryon and in the meson & 3 & $41^{+12}_{-3}$ & -- & $71^{+42}_{-3}$ & -- \\
\midrule
\multicolumn{7}{@{}l}{\emph{No light quarks}} \\
$\Omega^*_{QQQ}(Q\bar Q)$ & -- & 5 & $[-15,\,8]^\dagger$ & $[126,\,139]^\dagger$ & $-18^{+5}_{-8}$ & $[183,\,228]^\dagger$ \\
\bottomrule
\end{tabular*}
\end{table*}

\begin{figure}[t]
\centering
\includegraphics[width=\columnwidth]{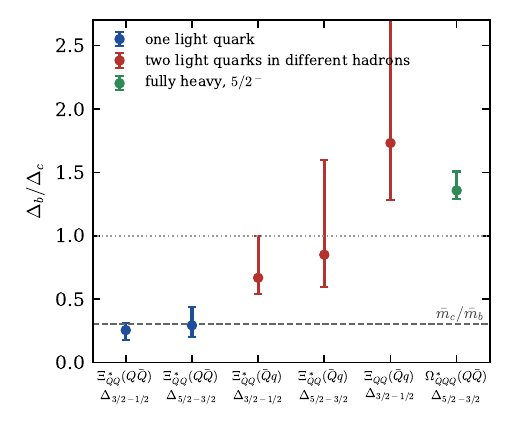}
\caption{Ratio of the splittings for bottom and charm, $\Delta_b/\Delta_c$, for the configurations with $|\Delta_c|\gtrsim40$~MeV, at the self-consistent thresholds with correlated uncertainties. For the $5/2^-$--$3/2^-$ splitting of the fully heavy systems the values at the opening of the $5/2^-$ window are used, $\sqrt{s_0}=10.6$~GeV for charm and $24.0$~GeV for bottom, with the variation of $\sqrt{s_0}$ by $\pm0.1$~GeV and, for charm, the input variations. The dashed line indicates $\barm_c/\barm_b$, the dotted line equal splittings.}
\label{fig:cb}
\end{figure}

The replacement of charm by bottom quarks separates these sources. The splittings of both flavors are compared in Table~\ref{tab:cb}, and the ratios $\Delta_b/\Delta_c$ for the configurations with $|\Delta_c|\gtrsim40$~MeV are shown in Fig.~\ref{fig:cb}. For the $\Xi^*_{QQ}(Q\bar Q)$ triplet, the only configuration with a single light quark and sizable splittings for charm, both splittings decrease from charm to bottom, by factors of $0.18$--$0.31$ and $0.20$--$0.43$, compatible with $\barm_c/\barm_b=0.30$. The configurations that are nearly degenerate for charm, $\Sigma^{(*)}_Q(Q\bar Q)$, $\Xi_{QQ}(Q\bar Q)$, and $\Omega^*_{QQQ}(\bar Qq)$, remain degenerate within a few MeV for bottom; for $\Omega^*_{QQQ}(\bar Qq)$, whose correlators contain no quark exchange, this is the degeneracy of a noninteracting system (Sec.~\ref{sec:conn}). For the configurations with two light quarks in different hadrons, $\Xi^*_{QQ}(\bar Qq)$ and $\Xi_{QQ}(\bar Qq)$, the splittings are not reduced: the ratios lie between about $0.55$ and $1.6$ for the triplet and between about $1.3$ and $3$ for the doublet. At a common threshold placed at the mean of the self-consistent thresholds of the members, the same pattern is obtained, with ratios of $0.29$ and $0.34$ for $\Xi^*_{QQ}(Q\bar Q)$, $0.8$ and $1.0$ for $\Xi^*_{QQ}(\bar Qq)$, and $2.0$ for $\Xi_{QQ}(\bar Qq)$.

This pattern can be understood from heavy-quark spin symmetry. In the heavy-quark limit the total spin $j_\ell$ of the light degrees of freedom is conserved separately from the heavy-quark spins. If the light degrees of freedom of a configuration consist of a single light quark, $j_\ell=1/2$, or of the light diquark of $\Sigma_Q$, $j_\ell=1$, all members of the configuration share the same $j_\ell$. Their mass differences then arise from the coupling of the heavy-quark spins and vanish as $1/m_Q$, which is what we observe for $\Xi^*_{QQ}(Q\bar Q)$ and, at the level of a few MeV, for the degenerate configurations. If the light degrees of freedom consist of two light quarks in different hadrons, the members of a configuration with different $J$ contain the light-spin states $j_\ell=0$ and $1$ with different weights. The energy difference between these light-spin states is governed by the interaction between the light quarks and does not vanish in the heavy-quark limit, consistent with the unsuppressed splittings of $\Xi^{(*)}_{QQ}(\bar Qq)$. As shown in Sec.~\ref{sec:conn}, this distinction holds separately for the two-hadron and the quark-exchange contributions to the splittings. The light-spin content can be made explicit by recoupling the spins of the four constituents of $\Xi^{(*)}_{QQ}(\bar Qq)$ into the heavy-quark spin and $j_\ell$. The probability of $j_\ell=1$ is $1/3$, $7/12$, and $1$ for the $J=1/2$, $3/2$, and $5/2$ members of $\Xi^*_{QQ}(\bar Qq)$, and $11/12$ and $2/3$ for the $J=1/2$ and $3/2$ members of $\Xi_{QQ}(\bar Qq)$. A single spin-dependent interaction between the two light quarks, evaluated with these weights, would order the triplet monotonically in $J$, with $\Delta_{3/2-1/2}/\Delta_{5/2-3/2}=3/5$. In both $\Xi^*$ triplets and for both flavors, however, the $3/2^-$ member lies lowest. The depression of this member is therefore not accounted for by the light-spin content of a single configuration and points to mixing with other configurations of the same $J^P$, in line with the overlap of the axial-type current with the $BV$ currents in Table~\ref{tab:overlap} and with the reduction of the $3/2^-$--$1/2^-$ splitting of $\Xi^*_{cc}J/\psi$ from $-94$ to $-42$~MeV when this overlap is removed. The size of the splittings follows the light degrees of freedom nevertheless: for $\Xi^*_{QQ}(Q\bar Q)$ they decrease by about $\barm_c/\barm_b$ for bottom, whereas for $\Xi^*_{QQ}(\bar Qq)$ they persist.

The fully heavy systems contain no light degrees of freedom, and their $1/2^-$ and $3/2^-$ states are degenerate within about $20$~MeV for both flavors. The $5/2^-$ state behaves differently. For both flavors its working region opens more than $1$~GeV above those of the other two channels, and its splitting from the $3/2^-$ state does not decrease from charm to bottom; at the opening of the respective windows it amounts to about $168$~MeV for charm and $228$~MeV for bottom. A hyperfine splitting would decrease from charm to bottom; in heavy quarkonia, for instance, the splitting between the vector and pseudoscalar ground states decreases from about $113$~MeV to about $62$~MeV~\cite{ParticleDataGroup:2024cfk}. The observed behavior is instead consistent with the interpretation of Sec.~\ref{sec:res5}: the spin-$5/2$ state cannot be formed in the S-wave configuration of four identical heavy quarks, and the $5/2^-$ current couples to configurations with internal relative motion, whose excitation energy is not suppressed by the heavy-quark mass. The decomposition of Sec.~\ref{sec:conn} shows directly that the splitting is generated by the exchange terms between the identical heavy quarks of the baryon and the meson. The interpretation of the fully heavy entries is nevertheless limited by the large thresholds at which they are obtained, $\sqrt{s_0}-m\approx1$~GeV for the $5/2^-$ states, where excited states and the continuum may contribute to the pole side. The values of $\Delta_{5/2-3/2}$ for $N_h=3$ and $4$ are further affected by the shared $5/2^-$ state, which we have analyzed for charm only.

\subsection{Two-hadron and quark-exchange contributions}\label{sec:conn}

The correlators of baryon--meson currents contain contributions in which the baryon and the meson propagate independently. In the Wick expansion these are the terms in which every propagator connects two fields of the same hadron, one at $x$ and one at $0$. We call these terms factorizable and the remaining ones, in which at least one quark is exchanged between the baryon and the meson, connected, and write
\begin{equation}
\Pi(M^2,s_0)=\Pi^{\rm fact}(M^2,s_0)+\Pi^{\rm conn}(M^2,s_0).
\label{eq:split}
\end{equation}
At the order considered here the factorizable part equals the product of the baryon and meson correlators in coordinate space, since a condensate whose gluons are attached to two different color-singlet hadrons vanishes; for $\Pi^{22}$ of $cccc\bar c$ this was verified numerically in Sec.~\ref{sec:res5}. The factorizable part therefore describes a noninteracting baryon and meson, including their two-hadron states, and the connected part contains the exchange of quarks between them. By quark--hadron duality the factorizable part corresponds to the two-hadron continuum starting at $(m_B+m_M)^2$, which is not contained explicitly in the hadronic representation of Eq.~(\ref{eq:sumrule}); in the finite-width test of Sec.~\ref{sec:width} part of this strength is attributed to the width of a single state. Gluon exchange between the two hadrons enters only at order $\alpha_s$ and is not included, so that ``connected'' refers to quark exchange only. Both parts are obtained from the same canonical Borel terms by restricting the sum over Wick contractions, and the decomposition is exact; for $cc\bar cqq$, recomputing two full correlators in this way reproduces all their terms.

If the baryon and the meson have no quark flavor in common, there are no connected terms. This is the case for the $\Omega^*_{QQQ}(\bar Qq)$ configurations, i.e., for $J_4$, $J_8$, $J_9$, and $J_{11}$ of $ccc\bar cq$ and $bbb\bar bq$, whose correlators are exactly factorizable at this order. For these currents the sum rules cannot distinguish a bound state from the $\Omega^*_{QQQ}\bar D^{(*)}$ or $\Omega^*_{QQQ}B^{(*)}$ two-hadron states. This applies in particular to the $3/2^-$ and $5/2^-$ ground states of $ccc\bar cq$, which are obtained from $J_8$ and $J_{11}$. The degeneracy of the $\Omega^*_{QQQ}(\bar Qq)$ triplet within $5$~MeV is that of a noninteracting system; it shows that the sum rules do not generate spurious splittings in this configuration.

In the other channels of $N_h=3$ and 4 the factorizable part dominates the correlators as well. At the self-consistent thresholds the ratio $\Pi^{\rm conn}/\Pi$ lies between $-2.0$ and $0.25$, and the masses obtained from $\Pi^{\rm fact}$ alone differ from those of Tables~\ref{tab:diag} and \ref{tab:diagb} by at most $83$~MeV. To determine the contribution of the quark exchange to a splitting we define $\Delta^{\rm conn}=\Delta-\Delta^{\rm fact}$, where $\Delta^{\rm fact}$ is obtained from $\Pi^{\rm fact}$ either (i) at the thresholds and Borel masses of the full analysis or (ii) at the self-consistent solutions of $\Pi^{\rm fact}$ itself, with the same criteria. The two definitions differ in how the change of the working regions is attributed and indicate the ambiguity of the decomposition. The results are collected in Table~\ref{tab:conn}.

\begin{table*}[t]
\caption{Connected (quark-exchange) parts $\Delta^{\rm conn}=\Delta-\Delta^{\rm fact}$ (MeV) of the splittings within configurations, with $\Delta^{\rm fact}$ obtained from the factorizable correlators of Eq.~(\ref{eq:split}) (i) at the self-consistent thresholds and Borel masses of the full correlators, with correlated uncertainties, and (ii) at their own self-consistent solutions. The second column gives the quark exchanged between the baryon and the meson. The $\Omega^*_{QQQ}(\bar Qq)$ configurations have no connected part. $^\dagger$Ranges at the common thresholds $\sqrt{s_0}=10.8$--$11.6$~GeV for charm and $24.0$--$24.2$~GeV for bottom, definition (i) only. A dash indicates that no self-consistent solution exists.}
\label{tab:conn}
\renewcommand{\arraystretch}{1.3}
\begin{tabular*}{\textwidth}{@{\extracolsep{\fill}}lllcccc@{}}
\toprule
 & & & \multicolumn{2}{c}{$Q=c$} & \multicolumn{2}{c}{$Q=b$} \\
\cmidrule(lr){4-5}\cmidrule(l){6-7}
Configuration & Exchanged & & (i) & (ii) & (i) & (ii) \\
\midrule
$\Sigma^*_Q(Q\bar Q)$ & $Q$ & $\Delta_{3/2-1/2}$ & $9^{+3}_{-2}$ & 24 & -- & -- \\
 &  & $\Delta_{5/2-3/2}$ & $-13^{+2}_{-3}$ & $-28$ & -- & -- \\
$\Sigma_Q(Q\bar Q)$ & $Q$ & $\Delta_{3/2-1/2}$ & $1^{+1}_{-0}$ & 2 & -- & -- \\
$\Xi^*_{QQ}(Q\bar Q)$ & $Q$ & $\Delta_{3/2-1/2}$ & $-15^{+4}_{-10}$ & $-32$ & $-5^{+2}_{-5}$ & $-7$ \\
 &  & $\Delta_{5/2-3/2}$ & $19^{+6}_{-2}$ & 34 & $5^{+5}_{-1}$ & 7 \\
$\Xi_{QQ}(Q\bar Q)$ & $Q$ & $\Delta_{3/2-1/2}$ & $8^{+1}_{-1}$ & 9 & $2^{+1}_{-0}$ & 5 \\
$\Xi^*_{QQ}(\bar Qq)$ & $q$ & $\Delta_{3/2-1/2}$ & $-57^{+17}_{-33}$ & $-125$ & $-47^{+17}_{-37}$ & $-79$ \\
 &  & $\Delta_{5/2-3/2}$ & $-2^{+7}_{-3}$ & $-3$ & $15^{+27}_{-10}$ & 23 \\
$\Xi_{QQ}(\bar Qq)$ & $q$ & $\Delta_{3/2-1/2}$ & $10^{+11}_{-4}$ & 24 & $27^{+29}_{-11}$ & 45 \\
$\Omega^*_{QQQ}(Q\bar Q)^\dagger$ & $Q$ & $\Delta_{3/2-1/2}$ & $[1,\,4]$ & -- & 0 & -- \\
 & & $\Delta_{5/2-3/2}$ & $[268,\,289]$ & -- & $[288,\,317]$ & -- \\
\bottomrule
\end{tabular*}
\end{table*}

For the configurations with $|\Delta|\gtrsim40$~MeV the factorizable part accounts for $60$--$100\%$ of the splittings with definition (i) and for $35$--$100\%$ with definition (ii). The distinction between the two classes of configurations of Sec.~\ref{sec:multiplets} is present in both parts. For $\Xi^*_{QQ}(Q\bar Q)$, where a heavy quark is exchanged, the connected parts of both splittings decrease from charm to bottom by factors of $0.28$ and $0.33$ with definition (i) and $0.21$ with definition (ii), and the factorizable parts by $0.24$--$0.32$. For the configurations in which a light quark is exchanged, the connected part of $\Delta_{3/2-1/2}$ changes by factors of $0.63$--$0.83$ for $\Xi^*_{QQ}(\bar Qq)$ and $1.9$--$2.7$ for $\Xi_{QQ}(\bar Qq)$, and the factorizable parts by $0.61$--$0.74$ and $1.4$--$1.5$; the connected part of $\Delta_{5/2-3/2}$ of $\Xi^*_{QQ}(\bar Qq)$ is compatible with zero for charm and amounts to $15$--$23$~MeV, with a large uncertainty, for bottom. In the heavy-quark limit the exchange of a heavy quark between the hadrons can depend on $J$ only through the heavy-quark spins and is expected to be suppressed by $1/m_Q$, whereas the exchange of a light quark carries the light spin, as found here. The depression of the $3/2^-$ member of $\Xi^*_{QQ}(\bar Qq)$ is already present in the factorizable part, which gives $-150$ and $-81$~MeV for charm with the two definitions. It thus arises from the coupling of the axial-type current to the two-hadron states of other configurations of the same $J^P$, in line with the interpretation of Sec.~\ref{sec:multiplets}.

The fully heavy systems behave differently. In their $1/2^-$ and $3/2^-$ channels $\Pi^{\rm conn}/\Pi$ amounts to $0.25$ and $-0.17$ to $-0.20$, and the connected terms change $\Delta_{3/2-1/2}$ by at most $4$~MeV. In the $5/2^-$ channel the factorizable and connected parts nearly cancel: at the common thresholds of Table~\ref{tab:conn}, $\Pi^{\rm conn}/\Pi$ lies between $-24$ and $-18$ for charm and between $-366$ and $-335$ for bottom. The factorizable part alone would place the $5/2^-$ state $134$--$159$~MeV (charm) and $36$--$95$~MeV (bottom) below the $3/2^-$ state. The connected terms, which antisymmetrize the heavy quark of the meson with those of the baryon, raise it by $268$--$289$ and $288$--$317$~MeV, an amount that does not decrease from charm to bottom. The large $5/2^-$--$3/2^-$ splitting of the fully heavy systems is thus generated by the exchange terms, as expected if it originates from the Pauli principle (Sec.~\ref{sec:res5}), and the near cancellation accounts for the high thresholds at which the $5/2^-$ working regions open.

\subsection{Comparison with other approaches}\label{sec:comparison}

Predictions for $cccc\bar c$ and $bbbb\bar b$ from other approaches are collected in Table~\ref{tab:lit}, with all values taken from the original works. The QCD sum rule analyses of Refs.~\cite{Zhang:2020vpz,Azizi:2024ito} employ the $\Omega^*_{ccc}\eta_c$ current, which coincides with our $J_2$, and extract the mass from the coefficient of $g_{\mu\nu}\pslash$ or $g_{\mu\nu}$ without parity projection. Reference~\cite{Zhang:2020vpz} includes $\langle g_s^2G^2\rangle$ and $\langle g_s^3G^3\rangle$ and works at $\sqrt{s_0}=7.7$--$8.2$~GeV and $M^2=3.5$--$4.8$~GeV$^2$, while Ref.~\cite{Azizi:2024ito} includes the gluon condensate with $\langle\alpha_sG^2/\pi\rangle=0.012$~GeV$^4$ and works at $s_0=60$--$64$~GeV$^2$ and $M^2=8$--$10$~GeV$^2$. Reference~\cite{Wang:2021xao} uses a diquark--diquark--antiquark current for $1/2^-$. With the parity projection, the bound ${\rm PC}\geq40\%$, and the convergence conditions of Sec.~\ref{sec:numerics}, our working regions require $\sqrt{s_0}\gtrsim8.9$~GeV, where the masses follow the threshold; our absolute $cccc\bar c$ masses are therefore not compared with these predictions, and the comparison is restricted to the splittings.

\begin{table*}[t]
\caption{Masses (MeV) of $cccc\bar c$ and $bbbb\bar b$ states from other approaches, taken from the original works, and the corresponding splittings $\Delta_{3/2-1/2}$ and $\Delta_{5/2-3/2}$ (MeV). For Ref.~\cite{Yang:2022bfu} the lowest resonance in each channel is quoted; for Ref.~\cite{Gordillo:2024sem} the total spin is $1/2$ and the three values refer to different cluster structures; for Ref.~\cite{Alonso-Valero:2024jim} the meson--baryon basis including the hyperfine term is quoted; for Ref.~\cite{An:2020jix} the masses in the modified scheme are quoted. For this work, the $bbbb\bar b$ masses are those of the $\Omega^*_{bbb}\Upsilon$ currents $J_1$ and $J_3$ at their self-consistent thresholds; the splittings are given as in Table~\ref{tab:cb}. A dash indicates that no value is available and ``n.b.'' that no bound state is found. Abbreviations: QCDSR, QCD sum rules; CMI, chromomagnetic interaction; ChQM, chiral quark model; QDCSM, quark delocalization color screening model; GR, G\"ursey--Radicati.}
\label{tab:lit}
\renewcommand{\arraystretch}{1.2}\setlength{\tabcolsep}{1.5pt}\footnotesize
\begin{tabular*}{\textwidth}{@{\extracolsep{\fill}}llccccc@{}}
\toprule
Approach & Ref. & $1/2^-$ & $3/2^-$ & $5/2^-$ & $\Delta_{3/2-1/2}$ & $\Delta_{5/2-3/2}$ \\
\midrule
\multicolumn{7}{@{}l}{$cccc\bar c$} \\
QCDSR, $\Omega^*_{ccc}\eta_c$ & \cite{Zhang:2020vpz} & -- & $7410^{+270}_{-310}$ & -- & -- & -- \\
QCDSR, $\Omega^*_{ccc}\eta_c$ & \cite{Azizi:2024ito} & -- & $7628\pm112$ & -- & -- & -- \\
QCDSR, $(cc)(cc)\bar c$ & \cite{Wang:2021xao} & $7930\pm150$ & -- & -- & -- & -- \\
CMI model & \cite{An:2020jix} & 7949 & 7864 & -- & $-85$ & -- \\
MIT bag model & \cite{Zhang:2023hmg} & 8262 & 8229 & -- & $-33$ & -- \\
Potential model & \cite{Liang:2024met} & 8242 & 8210 & -- & $-32$ & -- \\
Lattice-inspired model & \cite{Yang:2022bfu} & 8098 & 8095 & 8137 & $-3$ & 42 \\
ChQM and QDCSM & \cite{Yan:2021glh} & 7891.9--7892.7 & n.b. & n.b. & -- & -- \\
Diffusion Monte Carlo & \cite{Gordillo:2024sem} & 7923, 8021, 8194 & -- & -- & -- & -- \\
Hartree--Fock & \cite{Alonso-Valero:2024jim} & -- & -- & 7860(200) & -- & -- \\
Effective-mass scheme & \cite{Rashmi:2024ako} & 8537.4 & 8547.4 & 8562.5 & 10 & 15 \\
Extended GR formula & \cite{Sharma:2025grd} & $8356.85\pm91.01$ & $8425.85\pm91.12$ & $8540.85\pm91.61$ & 69 & 115 \\
Extended GR formula & \cite{Rostami:2026jyz} & -- & 7848 & -- & -- & -- \\
This work (QCDSR) & & -- & -- & -- & $[-15,\,8]$ & $[126,\,139]$ \\
\midrule
\multicolumn{7}{@{}l}{$bbbb\bar b$} \\
QCDSR, $\Omega^*_{bbb}\eta_b$ & \cite{Zhang:2020vpz} & -- & $21600^{+730}_{-220}$ & -- & -- & -- \\
QCDSR, $\Omega^*_{bbb}\eta_b$ & \cite{Azizi:2024ito} & -- & $21982\pm144$ & -- & -- & -- \\
QCDSR, $(bb)(bb)\bar b$ & \cite{Wang:2021xao} & $23910\pm150$ & -- & -- & -- & -- \\
CMI model & \cite{An:2020jix} & 23821 & 23775 & -- & $-46$ & -- \\
MIT bag model & \cite{Zhang:2023hmg} & 24770 & 24761 & -- & $-9$ & -- \\
Potential model & \cite{Liang:2024met} & 24251 & 24235 & -- & $-16$ & -- \\
Lattice-inspired model & \cite{Yang:2022bfu} & 24026 & 24059 & 24076 & 33 & 17 \\
ChQM and QDCSM & \cite{Yan:2021glh} & 23810.1--23813.8 & 23748.2--23752.3 & n.b. & $-62$ & -- \\
Hartree--Fock & \cite{Alonso-Valero:2024jim} & -- & -- & 23830(150) & -- & -- \\
Effective-mass scheme & \cite{Rashmi:2024ako} & 25212.8 & 25214.8 & 25216.3 & 2 & 2 \\
Extended GR formula & \cite{Sharma:2025grd} & $25683.90\pm172.00$ & $25752.90\pm172.06$ & $25867.90\pm172.32$ & 69 & 115 \\
Extended GR formula & \cite{Rostami:2026jyz} & -- & 24444 & -- & -- & -- \\
This work (QCDSR) & & $22350^{+104}_{-96}$ & $22333^{+106}_{-100}$ & -- & $-18^{+5}_{-8}$ & $[183,\,228]$ \\
\bottomrule
\end{tabular*}
\end{table*}

For $cccc\bar c$ the model predictions place the S-wave states between about $7.86$ and $8.56$~GeV, a spread of roughly $700$~MeV, whereas the splittings are more consistent. Where both values are available, $\Delta(3/2^--1/2^-)$ ranges from $-85$ to $69$~MeV; our range of $-15$ to $8$~MeV lies within this interval and is closest to Refs.~\cite{Yang:2022bfu,Rashmi:2024ako}. A $5/2^-$ state is reported in Refs.~\cite{Yang:2022bfu,Alonso-Valero:2024jim,Rashmi:2024ako,Sharma:2025grd}; where a $3/2^-$ value is given in the same work, it lies $15$ to $115$~MeV below the $5/2^-$ state. Our value of $126$--$139$~MeV lies somewhat above this range and is closest to the estimate of Ref.~\cite{Sharma:2025grd}. The absence of an S-wave $5/2^-$ state in the compact models of Refs.~\cite{An:2020jix,Zhang:2023hmg,Liang:2024met} is consistent with the absence of a spin-$5/2$ component in the nonrelativistic part of our tensor current.

For $bbbb\bar b$ our sum rules admit self-consistent solutions in the $1/2^-$ and $3/2^-$ channels, so that absolute masses are available for comparison. The same $\Omega^*_{bbb}\eta_b$ current was used in Ref.~\cite{Zhang:2020vpz} at $\sqrt{s_0}=21.8$--$23.0$~GeV and in Ref.~\cite{Azizi:2024ito} at $s_0=485$--$500$~GeV$^2$ and $M^2=25$--$35$~GeV$^2$, without parity projection. Our masses of $22.35$ and $22.33$~GeV lie about $0.35$--$0.75$~GeV above these two sum rule results and $1.5$~GeV below that of the diquark--diquark--antiquark current of Ref.~\cite{Wang:2021xao}. The model predictions lie between $23.75$ and $25.87$~GeV, i.e., about $1.4$~GeV or more above our values, and scatter by about $2$~GeV among themselves. Our values refer to $\barm_b(\barm_b)$; with the one-loop pole mass discussed below they increase to $24.29$--$24.31$~GeV, within the range of the model predictions, so that the comparison of absolute masses is not conclusive. The splittings are again more consistent. Where both values are given, $\Delta(3/2^--1/2^-)$ ranges from $-62$ to $69$~MeV; our value of $-18^{+5}_{-8}$~MeV lies within this interval and is closest to Refs.~\cite{Liang:2024met,Zhang:2023hmg}. In the compact models that address both flavors, Refs.~\cite{An:2020jix,Zhang:2023hmg,Liang:2024met}, the magnitude of this splitting decreases from charm to bottom by factors of $0.27$--$0.54$, whereas the mass formula of Ref.~\cite{Sharma:2025grd} gives identical splittings for both flavors. A $5/2^-$ state of $bbbb\bar b$ together with a $3/2^-$ partner from the same work is available from Refs.~\cite{Yang:2022bfu,Rashmi:2024ako,Sharma:2025grd}, at $17$, $2$, and $115$~MeV above the $3/2^-$ state; our range of $183$--$228$~MeV lies above all three values. As for charm, the compact models of Refs.~\cite{An:2020jix,Zhang:2023hmg,Liang:2024met} find no S-wave $5/2^-$ state, and the quark-model analysis of Ref.~\cite{Yan:2021glh} finds no bound $5/2^-$ state.

The thresholds $m_B+m_M$ of the configurations are given in Tables~\ref{tab:diag} and \ref{tab:diagb}. They are obtained from the masses of Ref.~\cite{ParticleDataGroup:2024cfk} and, for the baryons that have not been observed, from the lattice results $m_{\Xi^*_{cc}}=3692(28)(21)$, $m_{\Xi_{bb}}=10143(30)(23)$, $m_{\Xi^*_{bb}}=10178(30)(24)$, $m_{\Omega_{ccc}}=4796(8)(18)$, and $m_{\Omega_{bbb}}=14366(9)(20)$~MeV of Ref.~\cite{Brown:2014ena}. For the $\Sigma_b$ configurations, which do not appear in Table~\ref{tab:diagb}, the thresholds are $15.209$ ($\Sigma_b\eta_b$), $15.229$ ($\Sigma^*_b\eta_b$), $15.271$ ($\Sigma_b\Upsilon$), and $15.291$~GeV ($\Sigma^*_b\Upsilon$), and for $cccc\bar c$ they are $7.780$ ($\Omega^*_{ccc}\eta_c$) and $7.893$~GeV ($\Omega^*_{ccc}J/\psi$). For the charm systems the extracted masses lie close to or above the lowest thresholds. The $1/2^-$ and $3/2^-$ ground states of $cc\bar cqq$ lie $0.08$--$0.14$~GeV above the $\Sigma_c\eta_c$, $\Xi_{cc}\bar D$, $\Sigma^*_c\eta_c$, and $\Sigma_cJ/\psi$ thresholds, i.e., by at most about 1.5 times their uncertainties, while the $5/2^-$ ground state lies $0.27$~GeV above the $\Sigma^*_cJ/\psi$ threshold. The $3/2^-$ ground state of $ccc\bar cq$, obtained from the exactly factorizable $\Omega^*_{ccc}\bar D$ current, is compatible with the $\Omega^*_{ccc}\bar D$ threshold, whereas the $1/2^-$ ground state lies about $0.17$~GeV above the $\Xi_{cc}J/\psi$ threshold and the members of the $\Omega^*_{ccc}\bar D^*$ triplet about $0.22$~GeV above the $\Omega^*_{ccc}\bar D^*$ threshold. These distances refer to $\barm_c(\barm_c)$; their dependence on the mass scheme is discussed below.

For the bottom systems the situation is different. All extracted masses lie below the lowest thresholds, by $0.43$--$0.62$~GeV for $bb\bar bqq$, $0.98$--$1.14$~GeV for $bbb\bar bq$, and $1.41$--$1.44$~GeV for $bbbb\bar b$; the two sum rule results with the $\Omega^*_{bbb}\eta_b$ current in Refs.~\cite{Zhang:2020vpz,Azizi:2024ito} lie even lower. The deficit grows with the number of bottom quarks, which suggests that it is related to the treatment of the heavy-quark mass rather than to binding. We use the $\overline{\rm MS}$ mass $\barm_b(\barm_b)$ in a leading-order perturbative term, where the choice of the mass scheme is not fixed at the accuracy of the calculation. To quantify this dependence we have repeated the analysis of the bottom systems with the one-loop pole mass $m_b=\barm_b\big[1+4\alpha_s(\barm_b)/(3\pi)\big]=4.58$~GeV, with $\alpha_s(\barm_b)=0.224$ from the running used throughout and all other inputs and criteria unchanged. All self-consistent solutions of Table~\ref{tab:diagb} persist. The masses increase by $1.19$--$1.20$, $1.58$--$1.59$, and $1.96$~GeV for $bb\bar bqq$, $bbb\bar bq$, and $bbbb\bar b$, close to $N_h(m_b-\barm_b)$, and then lie $0.58$--$0.76$, $0.45$--$0.60$, and $0.52$--$0.54$~GeV above the lowest thresholds. The position of the bottom states relative to the thresholds is thus determined by the mass scheme at this order. The splittings within configurations, in contrast, change by at most $5$~MeV, since the members of a configuration share the same heavy-quark content. The absolute masses of the bottom systems are therefore much less reliable than their splittings, which are the quantities used in Sec.~\ref{sec:multiplets}.

The same applies to the charm systems. With the one-loop pole mass $m_c=\barm_c\big[1+4\alpha_s(\barm_c)/(3\pi)\big]=1.48$~GeV, with $\alpha_s(\barm_c)=0.384$, all self-consistent solutions persist, and the masses increase by $0.60$--$0.68$~GeV for $cc\bar cqq$ and $0.79$--$0.83$~GeV for $ccc\bar cq$, again close to $N_h(m_c-\barm_c)$. The ground states then lie $0.72$--$1.03$~GeV above the thresholds of their channels instead of $0.04$--$0.27$~GeV, so that their position relative to the thresholds is not determined either. The splittings within configurations change by at most $21$~MeV, which is small compared with the shifts of the masses but larger than for bottom. The pattern discussed in Sec.~\ref{sec:multiplets} is not affected: with the pole masses for both flavors the ratios of the bottom and charm splittings are $0.25$ and $0.30$ for $\Xi^*_{QQ}(Q\bar Q)$, $0.77$ and $0.99$ for $\Xi^*_{QQ}(\bar Qq)$, and $1.7$ for $\Xi_{QQ}(\bar Qq)$, compared with $0.25$, $0.29$, $0.67$, $0.84$, and $1.7$ with the $\overline{\rm MS}$ masses.

\section{Summary}\label{sec:summary}

We have studied negative-parity pentaquarks with hidden heavy flavor along the chains $QQ\bar Qqq\to QQQ\bar Qq\to QQQQ\bar Q$ for $Q=c$ and $b$, with the complete set of S-wave baryon--meson currents and a common QCD sum rule framework based on the parity-projected sum rules, a pole-contribution bound of $40\%$, a bound on the highest computed dimension, Borel stability, and a self-consistent continuum threshold. For $cc\bar cqq$ and $ccc\bar cq$ all channels admit self-consistent solutions, and the ground-state masses are $5.579$, $5.630$, and $5.889$~GeV and $6.891$, $6.697$, and $7.025$~GeV for $J^P=1/2^-$, $3/2^-$, and $5/2^-$, with uncertainties of $60$--$170$~MeV dominated by $m_0^2$ and $s_0$. A Breit--Wigner width of up to $200$~MeV changes these masses by at most $69$~MeV and the splittings within configurations by at most $39$~MeV; since the correlators are dominated by two-hadron contributions (see below), this insensitivity does not by itself establish a resonance. For $cccc\bar c$ the gluon condensate moves the working regions to thresholds where no self-consistent solution exists. In the bottom systems the gluon condensate is much less important, and self-consistent solutions exist for all currents of $bbb\bar bq$, for the $\Xi_{bb}$ configurations of $bb\bar bqq$, and for the $1/2^-$ and $3/2^-$ channels of $bbbb\bar b$, with masses of $14.59$--$14.78$, $18.40$--$18.57$, and $22.33$--$22.35$~GeV. These masses lie $0.4$--$1.4$~GeV below the lowest thresholds, with a deficit that grows with the number of bottom quarks. With the one-loop pole masses instead of the $\overline{\rm MS}$ masses, the masses of both flavors increase by about $N_h(m_Q-\barm_Q)$ and lie above the thresholds, by $0.72$--$1.03$~GeV for the charm ground states and $0.45$--$0.76$~GeV for the bottom states, whereas the splittings within configurations change by at most $21$~MeV for charm and $5$~MeV for bottom, and the ratios of the bottom and charm splittings remain stable. The absolute masses are therefore less reliable than the splittings.

Since the ground states of different $J^P$ arise from different configurations, the spin structure is better characterized by the splittings within a configuration, which are also much less sensitive to the input parameters. The comparison of charm and bottom indicates that their size is governed mainly by the structure of the light degrees of freedom. Configurations whose light degrees of freedom have a unique spin, a single light quark or the light diquark of $\Sigma_Q$, are either nearly degenerate for both flavors or, as for $\Xi^*_{QQ}(Q\bar Q)$, have splittings that decrease from charm to bottom by a factor compatible with $\barm_c/\barm_b$. Configurations with two light quarks in different hadrons have splittings of $40$--$210$~MeV that are not reduced for bottom, as expected from heavy-quark spin symmetry when the members contain different light-spin components. In the fully heavy systems the $1/2^-$ and $3/2^-$ states are degenerate for both flavors, whereas the $5/2^-$ state requires a threshold more than $1$~GeV higher and lies $126$--$139$~MeV ($cccc\bar c$) and $183$--$228$~MeV ($bbbb\bar b$) above the $3/2^-$ state. Together with the absence of a spin-$5/2$ component in the nonrelativistic limit of the tensor current, this suggests that the $5/2^-$ state is excluded from the S-wave configuration by the Pauli principle rather than being a hyperfine partner of the other two states. These conclusions rest on the pure currents; the overlap of the $5/2^-$ currents of different configurations, analyzed here for charm, and the large thresholds of the fully heavy $5/2^-$ states limit the precision of the corresponding entries.

The decomposition of the correlators into factorizable two-hadron and connected quark-exchange contributions qualifies and supports these conclusions. The factorizable part dominates the correlators and accounts for most of the splittings, between about $35$ and $100\%$ depending on the definition of the decomposition. The correlators of configurations without a common quark flavor, $\Omega^*_{QQQ}(\bar Qq)$, are exactly factorizable at this order, so that the corresponding states, including the $3/2^-$ and $5/2^-$ ground states of $ccc\bar cq$, cannot be distinguished from two-hadron states. The distinction between configurations with a unique light spin and those with two light quarks in different hadrons is present in both parts: the quark-exchange contributions decrease from charm to bottom by factors of about $0.2$--$0.3$ when a heavy quark is exchanged and are not reduced when a light quark is exchanged. In the fully heavy $5/2^-$ channels the factorizable part alone would place the state below the $3/2^-$ one, and the exchange terms between the identical heavy quarks raise it by about $270$--$320$~MeV for both flavors, which supports the interpretation in terms of the Pauli principle.

\appendix

\section{Spectral densities of the $\Xi_{cc}\bar D$ correlator}\label{app:rho}

As an example of the structure of the operator product expansion we give the perturbative and quark-condensate contributions to the spectral densities of the $\Xi_{cc}\bar D$ current $J_4$ of $cc\bar cqq$, which yields the $1/2^-$ ground state. The three heavy lines are described by Feynman parameters $y_1$, $y_2$, $y_3$ with $\int d\mu=\int_0^1dy_1\,dy_2\,dy_3\,\delta(1-y_1-y_2-y_3)$, and we use $e_2=y_1y_2+y_1y_3+y_2y_3$, $e_3=y_1y_2y_3$, and $\sigma=e_2/e_3$. The parameter $t$ arises from the Fourier transform of the two light-quark lines. With
\begin{equation}
\mathcal K_n(s;a)=\frac{(s-a)^n}{n!}\,\theta(s-a)
\end{equation}
and $\mathcal K_n\equiv\mathcal K_n(s;m_c^2\sigma/t)$ unless indicated otherwise, the densities of the structures $\pslash$ and $1$ read
\begin{align}
\rho_A^{(0)}(s)&=\frac{1}{\pi^8}\int d\mu\int_0^1dt\left[\frac{33}{1024}\,e_3\,t^5(1-t)\,\mathcal K_5+\frac{11}{4096}\,m_c^2\,t^3(1-t)^2\,\mathcal K_4\right],\\
\rho_B^{(0)}(s)&=-\frac{m_c}{\pi^8}\int d\mu\int_0^1dt\left[\frac{e_2}{2048}\,t^4(1-t)\,\mathcal K_5+\frac{3m_c^2}{2048}\,t^2(1-t)^2\,\mathcal K_4\right],\\
\rho_A^{(3)}(s)&=-\frac{5m_c\langle\bar qq\rangle}{\pi^6}\int d\mu\int_0^1dt\left[\frac{e_2}{384}\,t^4\,\mathcal K_3+\frac{m_c^2}{256}\,t^2(1-t)\,\mathcal K_2\right],\\
\rho_B^{(3)}(s)&=-\frac{5\langle\bar qq\rangle}{\pi^6}\left[\frac{m_c^2}{384}\int d\mu\int_0^1dt\,t^3\,\mathcal K_3+\frac{1}{256}\int d\mu\;e_3\,\mathcal K_4(s;m_c^2\sigma)\right].
\end{align}
Since the remaining factors depend on the $y_i$ only through $\sigma$, the monomials in the $y_i$ have been symmetrized. The densities enter Eq.~(\ref{eq:sumrule}) in the combination $\sqrt{s}\,\rho_A(s)+\rho_B(s)$. As a check of the normalization, with $\barm_c=1.2730$~GeV and $\langle\bar qq\rangle(2~{\rm GeV})=-(0.2791~{\rm GeV})^3$ the perturbative and quark-condensate contributions to $\Pi(M^2,s_0)$ at $\sqrt{s_0}=6.08$~GeV and $M^2=3.73$~GeV$^2$ amount to $9.98\times10^{-9}$ and $4.99\times10^{-8}$~GeV$^{13}$.

\bibliographystyle{elsarticle-num}
\bibliography{Nh_pentaquarks}

\end{document}